\documentclass[11pt,aps,prd,preprint,preprintnumbers,showpacs,superscriptaddress,floatfix]{revtex4-1}
\usepackage{graphicx}
\usepackage{epsfig}
\usepackage{dcolumn}
\usepackage{bm}
\usepackage{subfigure}
\usepackage{amsmath}
\usepackage{amssymb}
\usepackage{bbm}
\usepackage{multirow}
\usepackage{setspace}
\usepackage{xcolor}
\usepackage[colorlinks,
            linkbordercolor={0 0 0},
            pdfborder={0 0 0},
            linkcolor=blue,
            citecolor=blue,
            urlcolor=blue,
            breaklinks=true]{hyperref}

\allowdisplaybreaks

\newlength{\x}
\newlength{\y}
\newlength{\z}
\phantomsection
\begin{document}
%\preprint{IMSc/2016/10/06}

\title{ $Z_3$ meta-stable states in PNJL model.}

\author{Minati Biswal}%
\email{minati.b@iopb.res.in}%
\affiliation{Institute of Physics, Bhubaneswar, 751005, India}%

\author{Sanatan Digal}%
\email{digal@imsc.res.in}%
\affiliation{The Institute of Mathematical Sciences, Chennai, 600113, India}%

\author{P. S. Saumia}%
\email{saumia@theor.jinr.ru}%
\affiliation{The Institute of Mathematical Sciences, Chennai, 600113, India}%
\affiliation{Bogoliubov Laboratory of Theoretical Physics, JINR, 141980 Dubna, Russia}%
%1.905*0.190= 0.36195  (meta-stable state appear)

\begin{abstract}
We study the $Z_3$ meta-stable states in the Polyakov loop Nambu-Jona-Lasinio (PNJL) model. 
These states exist for temperatures above $T_m \sim 194$ MeV and can decay
via bubble nucleation. We numerically solve the bounce equation to compute 
the nucleation rate. We speculate that, in the context of heavy-ion collisions,
the likely scenario for the decay of the meta-stable states is via spinodal
decomposition. 
	
\end{abstract}

%\pacs{11.10.Wx,11.15.Ha,11.15.-q}
\maketitle
\section{Introduction} 
\label{sec:intro}  
\noindent
In pure SU(N) gauge theories, energy density increases sharply across the critical 
temperature ($T_c$). It is believed that this is due to deconfinement of the constituents 
(gluons) of low energy excitations of the theory. This transition from confined to 
deconfined state of gluons, known as confinement-deconfinement (CD) transition, has been extensively 
studied in the literature~\cite{Kuti:1980gh, Celik:1983wz, Susskind:1979up, Engels:1980ty, 
Karsch:2001cy}. The CD phase transition is found to be second order for $N=2$ 
~\cite{Damgaard:1987wh, Engels:1988ph, Christensen:1990qs, Damgaard:1994np, Engels:1994xj, 
Engels:1998nv} and, first order for $N\ge 3$~\cite{Boyd:1995zg, Boyd:1996bx}. The Polyakov 
loop, which transforms as a $Z_N$ spin,  plays the role of an order parameter. It is real valued 
for $N=2$ and complex for $N>2$. Above the critical temperature, in deconfined phase, it acquires 
a non-zero expectation value, spontaneously breaking the $Z_N$ symmetry. This leads to $N$ 
degenerate vacua. This non-trivial nature of the deconfined state allows for the existence 
of topological defects such as, domain walls for $N=2$, and domain walls connected by
strings for $N>2$~\cite{Balachandran:2001qn, Layek:2005fn, Gupta:2010pp}.

\vskip 0.1cm
In a realistic theory such as quantum chromo dynamics (QCD), there are fermions (quarks) in 
the fundamental representation. The presence of these fermions lead to explicit breaking of 
the $Z_N$ symmetry. The strength of the explicit symmetry breaking depends on the quark masses 
as well as the number of quark flavors~\cite{Green:1983sd, Oevers:1997yf, Karsch:2000zv, 
Belyaev:1991np, Deka:2010bc}. It affects the nature of the CD transition~\cite{Green:1983sd, 
Karsch:2000zv} as well as the transition temperature. For large explicit symmetry breaking, the 
CD transition turns into a cross-over while the transition temperature tends to decrease. 
Furthermore, there are no $N$ degenerate vacua in the deconfined phase. Out of the previous $N$ 
vacua, all but one becomes the ground state. With explicit symmetry breaking, the topological 
defects can still exist, but far above $T_c$ and most of them are time dependent 
(non-static)~\cite{Gupta:2011ag}.

\vskip 0.1cm
The explicit breaking of the $Z_N$ symmetry due to matter fields has been studied by 
calculating the the partition function or the effective potential of the Polyakov loop, 
when the gauge and matter field fluctuations are small ~\cite{Gross:1980br,
Weiss:1980rj, Weiss:1981ev, Guo:2018scp}. These perturbative calculations are reliable 
for high temperatures ($T >> T_c$), when the gauge coupling is 
expected to be small. Calculations which include fluctuations up to second order (one loop) 
show the presence of meta-stable states~\cite{Gross:1980br, Weiss:1981ev, Belyaev:1991np, Dixit:1991et}.               
%{\bf nathan weiss}, 
These states have been studied extensively in the context of cosmology. In the early Universe, they 
are found to be long lived, and can leave observable imprints while decaying nucleation of bubbles of true vacuum 
as in a first order transition~\cite{Dixit:1991et, Ignatius:1991nk}. However, the number 
of effective quark flavors is larger than 3, in which case, the free energies of the 
meta-stable states, at one loop, are positive and hence lead to negative pressure and 
entropy~\cite{Belyaev:1991np}. This problem doesn't arise in QCD near the critical temperature 
as the number of flavors is effectively $\leq 3$.

\vskip 0.1cm

The study of $Z_3$ meta-stable states for small temperatures, in particular near $T_c$ is important  
as they may affect the the evolution of quark-gluon plasma (QGP) in heavy-ion 
collision experiments. Near $T_c$ perturbative calculations are expected to break down due to 
large gauge coupling constant and fluctuations. There are very few studies of $Z_N$ symmetry using non-perturbative 
lattice QCD simulations. Lattice QCD results for 2 flavors show that out of the previously 3 degenerate 
vacua only one remains stable, while the other two become meta-stable states~\cite{Deka:2010bc}. The two 
meta-stable states are degenerate, related via $Z_2$ symmetry. Further, the meta-stability depends 
on the temperature, with the meta-stable states becoming unstable below 2$T_c$ ~\cite{Deka:2010bc}. 
In general, non-perturbative lattice simulations are essential for a quantitative estimate of the 
explicit symmetry breaking; however, the mean field approaches provide a qualitative understanding. 
In a recent study of $Z_3$ symmetry in the Polyakov loop quark meson (PQM) model~\cite{Mishra:2016ipq}, it 
is found that the meta-stable states exist above $~310$ MeV.

\vskip 0.1cm
In heavy-ion collisions, the initial conditions are far from equilibrium. The system quickly 
thermalizes in less than a $fm$ time. In such a scenario, it is possible that the whole or part 
of the system can get trapped in one of the meta-stable states~\cite{Ignatius:1991nk}. Also, if 
the system somehow thermalizes to a state of super hot hadron gas, which is a possibility at high 
baryon density, it will decay through bubble nucleation and some of the bubbles will have 
meta-stable cores.

\vskip 0.1cm
In the present work, the $Z_3$ meta-stable states are studied in the PNJL model at zero baryon
chemical potential. In this model, they exist above $T_m \sim 194$ MeV. If such a state exists, 
it can either become unstable (when temperature drops below $T_m$) or decay through nucleation of 
bubbles which grow in real time converting the meta-stable state to stable state. To compute the 
nucleation rate, the Euler-Lagrange equation for the bubble/bounce solution~\cite{Coleman:1977py, 
Callan:1977pt, Linde:1981zj, Gleiser:1991rf} is numerically integrated. The action as well as other 
properties of the bounce solution are found to depend strongly on the temperature. This study finds 
that the likely scenario for the evolution/decay of a meta-stable state in heavy ion collision is 
spinodal decomposition. This will lead to large oscillations of the Polyakov loop.  We mention here 
that in heavy-ion collisions the baryon chemical potential is small but non-zero. At finite chemical
potential the thermodynamic potential has an imaginary part. There are several papers that discuss 
 how to include the effect of non-zero $\mu$~\cite{Dumitru:2005ng, Rossner:2007ik}. With 
$\mu$, the  contribution of the fermions to the free energy will increase. Since the fermions break 
the $Z_N$ symmetry, we expect that finite $\mu$ will lead to more explicit breaking. Following 
Roessner et.al., for small $\mu$, we calculated the thermodynamic potential to the leading order, i.e, 
keeping only the real terms and found that $T_m$ increases slightly with $\mu$~\cite{Mishra:2016ipq}.
With increase in $T_m$ there is lesser time available for the nucleation of bubbles which enhances the likelihood of meta-stable
states becoming unstable.

\vskip 0.1cm
The paper is organized as follows. In section II, $Z_3$ symmetry in pure $SU(3)$ gauge theory 
is discussed. We briefly go through the explicit breaking of $Z_3$ symmetry in the PNJL model 
and compute the thermodynamic properties of the meta-stable states in section III. In section IV, 
we present the calculation of the bounce solution. In section V, we discuss the evolution of the 
meta-stable states in heavy-ion collisions and present our conclusions in section VI.

\smallskip

%%%%%%%%%%%%%%%%%%%%%%%%%%%%%%%%%%%%%%%%%%%%%%%%%%%%%%%%%%%%%%%%%%%%%%%%%%%%%%%%%%%%%%%%% 
\section{ $Z_3$ symmetry in pure gauge theory.}  
%%%%%%%%%%%%%%%%%%%%%%%%%%%%%%%%%%%%%%%%%%%%%%%%%%%%%%%%%%%%%%%%%%%%%%%%%%%%%%%%%%%%%%%%% 

In path integral formulation, gauge fields which are periodic in the temporal direction only contribute
to the partition function, i.e
\begin{equation}
A_\mu(\vec{x},0)=A_\mu(\vec{x},\beta),
\end{equation}
 where $\beta={1 \over T}$.
This boundary condition allows for the gauge transformations, $U(\vec{x},\tau)$, to be periodic up to 
a factor $z\in Z_N$, such as

\begin{equation}
U(\vec{x},0)=zU(\vec{x},\beta).
\label{uz}
\end{equation}
Though the partition function is invariant under the above gauge transformation, the Polyakov loop 
transforms as a $Z_N$ spin. 
%In this section, we discuss the deconfinement phase transition in pure SU(3) gauge 
%theory. We consider the potential for the Polyakov loop, which is consistent with the global $Z_3$ 
%symmetry~\cite{Dumitru:2000in, Dumitru:2001vc}. 
%The potential is obtained using the thermal 
%average of the Polyakov loop (L) which is a order parameter of the theory. 
The Polyakov loop is defined as
\begin{equation}
L(\vec{x}) = {1 \over 3} {\rm Tr}\left({\cal P}~{\rm exp} \left[ ig \int_0^{\beta} d\tau A_0 ({\vec{x}}, \tau) \right]\right)
\end{equation}
Here `${\cal P}$' denotes path order, `g' is the gauge coupling and $A_0 = A_0^a {\tau^a \over 2}$ is the 
temporal gauge field. Here $\tau_a$ are the Pauli matrices with `$a$' denoting the color indices. 
Under a $Z_3$ gauge transformation,~Eq. (\ref{uz}), the Polyakov loop, transforms as 
$L(\vec{x})\rightarrow zL(\vec{x})$. The thermal as well as volume average of the Polyakov loop $L(\vec{x})$,
\begin{equation}
L(T) = \left< {1 \over V} \int L(\vec{x}) d^3x\right>,
\end{equation}	
is related to the free energy $F_{\bar{Q}Q}(r)$ of a static (infinitely heavy) 
quark-anti-quark pair at infinite separation~\cite{Svetitsky:1985ye}.
\begin{equation}
|L(T)|^2= {\rm exp}\left[-\beta F_{\bar{Q}Q}(r=\infty)\right].
\end{equation}

In the following, we briefly describe the $Z_3$ symmetry in the effective potential for the Polyakov 
loop which describes the CD transition in pure SU(3) gauge theory~\cite{Pisarski:2000eq}. 
We consider the following Landau-Ginsburg effective potential for a complex field 
$\Phi$~\cite{Pisarski:2000eq, Dumitru:2000in, Dumitru:2001vc}. 
%(\textcolor{red}{ Pisarski's original model should be refereed here, others}),
\begin{equation}
	{\it{U}} ( {\bar{\Phi}}, \Phi , T ) = b_4 T^4 \left[ -{{b_2 (T)} \over 4} 
	({|\Phi|^2+{|\bar{\Phi}|^2}}) - {{b_3 } \over 6} (\Phi^3 +{\bar{\Phi}}^3 ) 
+ {{1} \over 16} (|\Phi|^2+{|\bar{\Phi}|^2} )^2 \right].
\label{polpot}
\end{equation}
Different forms of the effective potential in terms of the field $\Phi$ have been proposed
~\cite{Dumitru:2000in, Ratti:2005jh, Ratti:2006wg, Roessner:2006xn, Bhattacharyya:2010jd, Ghosh:2007wy}.
Across the critical temperature $T_0$, the Polyakov loop expectation value jumps discontinuously. 
The $Z_3$ symmetry and the first order nature of the CD transition require a cubic term in the 
effective potential. The factor $T^4$ takes care of the dimension of the effective 
potential~\cite{Pisarski:2000eq}. In the mean-field approximation the minimum (minima), $\Phi_{th}$, 
of the effective potential gives the thermal average of the Polyakov loop, $L(T)$. $\Phi_{th} = L(T)$. 
In this approximation the pressure $P$ is given by,
\begin{equation}
{\rm P} = -U(\bar{\Phi}_{th},\Phi_{th},T).
\end{equation}
The above 
effective potential with the following form of $b_2 (T)$
\begin{equation}
{b_2 (T)} = \left(1- 1.11 ~T_0 / T \right) \left(1 + 0.265~ T_0 / T \right)^2 
 \left(1 + 0.3 ~T_0 / T \right)^3 -0.487
\end{equation}
and the coefficients $b_3 = 2.0$ and $b_4 = 0.6016$ reproduces the pressure of the pure gauge 
theory computed from non-perturbative lattice method(s). Following~\cite{Dumitru:2000in}, 
for QCD, $b_4$ is rescaled as $b_4=0.6061 \times 47.5/16$. Here the coefficients are chosen such 
that the expectation value of 
the order parameter $\Phi$ approaches unity for $T \rightarrow \infty$. Hence, the fields 
and the coefficients in the above potential are rescaled as $\Phi \rightarrow \Phi /x$, 
$b_2 (T) \rightarrow b_2 (T) /x^2$, $b_3 \rightarrow b_3 /x$, $b_4 \rightarrow b_4 /x^4$, 
where  $x= b_3 /2 + {1 \over 2} \sqrt{b_3 ^2 +4b_2 (T = \infty)}$ for temperature $T \rightarrow \infty$.
Note that we have used a polynomial form of the Polyakov loop potential which is different from the 
more commonly used form as in~\cite{Ratti:2005jh}.  The $T_m$ for the latter case
  is higher than the initial temperature of QGP in heavy ion collisions which is not in agreement 
with lattice study~\cite{Deka:2010bc}. Also it is not clear that these models will be 
valid at such high temperatures.

%\textcolor {red}{The important difference between the two forms of the potentials is the 
%difference in barrier heights between the degenerate vacua.}  

\begin{figure}[h]
  \centering
  \subfigure[]
  {\rotatebox{0}{\includegraphics[width=0.40\hsize]
      {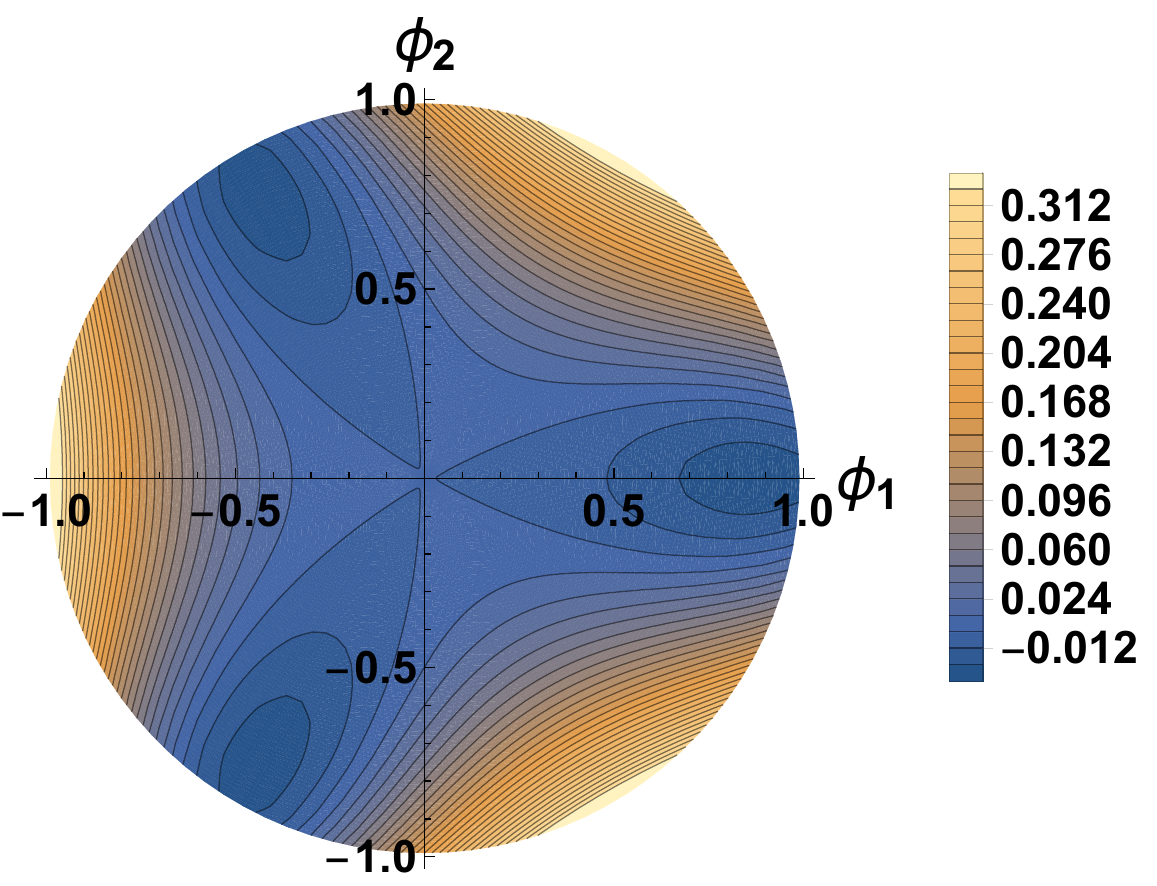}}
    \label{fig1a}
  }
    \subfigure[]
  {\rotatebox{0}{\includegraphics[width=0.40\hsize]
      {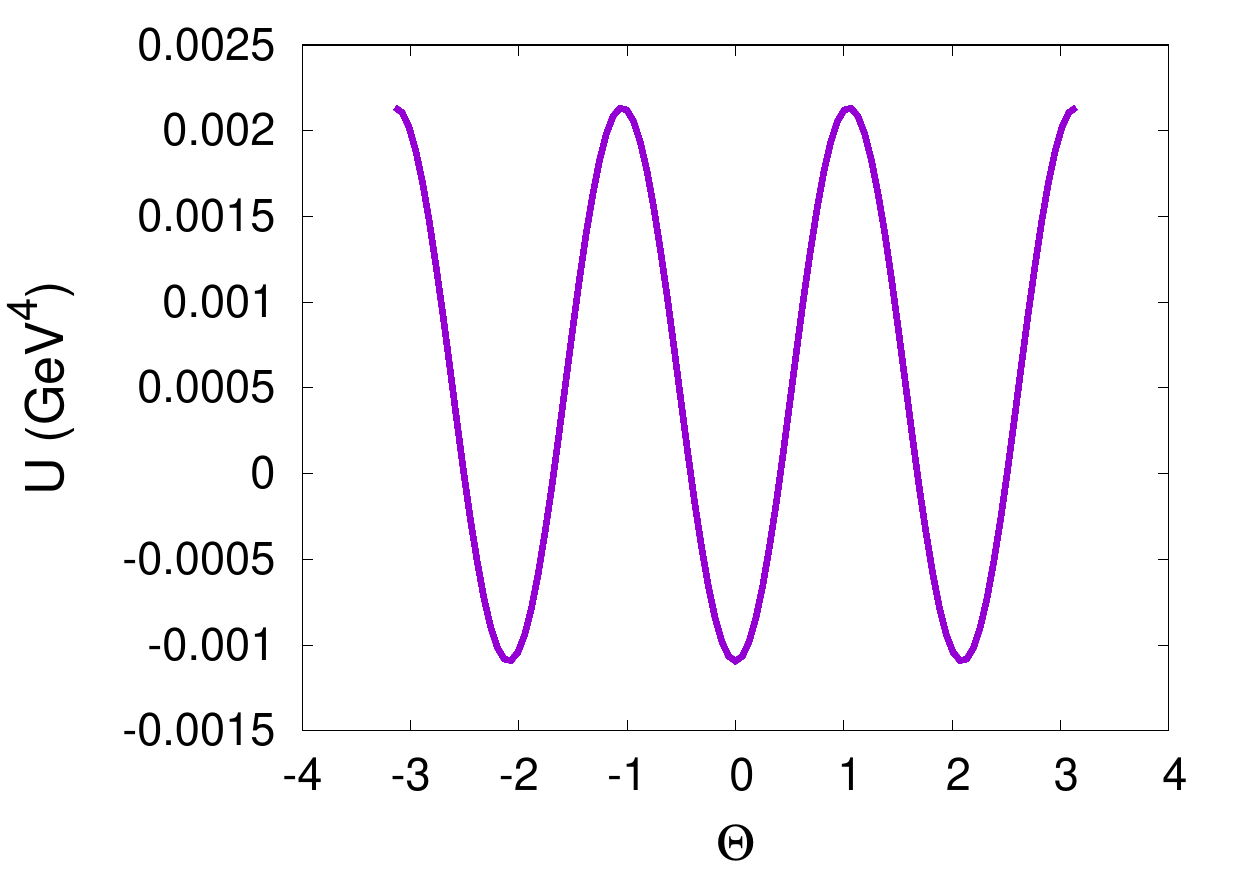}}
    \label{fig1b}
  }
\caption{(a) Contour plot for the Polyakov loop effective potential 
on the $\Phi_1$ - $\Phi_2$ plane at 1.3$T_0$, (b) Thermodynamic potential 
versus $\theta$ at 1.3$T_0$.}
\end{figure}

For $T>T_0$, there are three degenerate minima in the effective potential, which can be seen in 
Fig.~\ref{fig1a}, the contour plot of the effective potential in the complex 
$\Phi= (\Phi_1,\Phi_2)$ plane at $T=1.3T_0$. We also plot the variation of the potential along a 
circle going through the three minima, i.e $\Phi = |\Phi_{th}|e^{i\theta}$ in Fig.~\ref{fig1b}. 
The three degenerate vacua are situated at $\theta = 0$, ${{2 \pi} \over 3}$ and ${{4 \pi} \over 3}$, 
related by $Z_3$ rotation.

When dynamical quarks are included, the $Z_3$ symmetry is broken. 
While the pure gauge part of the action is $Z_3$ symmetric, the 
quark part of the action is not invariant under the $Z_3$ gauge transformations. 
This is because the gauge transformed quark fields are no longer anti periodic 
along the temporal direction. The non-trivial $Z_3$ gauge transformations can act 
only on the gauge fields. The situation is similar to the presence of an external 
(explicit breaking) field in spin systems. For example, in the presence of an external 
field the Ising model Hamiltonian has both $Z_2$ symmetric and broken terms. As the 
magnetization still describes the Ising transition, the field $\Phi$ too describes 
the CD transition~\cite{Engels:1993tp}. In the following section, we discuss the 
PNJL model which provides a prescription to include the effect of quarks on 
the $Z_3$ symmetry and the Polyakov loop effective potential.

%%%%%%%%%%%%%%%%%%%%%%%%%%%%%%%%%%%%%%%%%

\section{Meta-stable states in PNJL model}

The PNJL model is an extension of the Nambu-Jona-Lasinio (NJL) model. 
The NJL model is a phenomenological model formulated on the basis of the chiral symmetry 
of QCD and describes the dynamics of low energy excitations as well as the chiral 
transition~\cite{Nambu:1961tp, Nambu:1961fr, Klevansky:1992qe, Vogl:1991qt, Scavenius:2000qd, 
Hatsuda:1994pi, Buballa:2003qv, Nebauer:2001rb}. Since there are no gauge fields in this 
model, it can not describe the CD transition. The PNJL model attempts to include the gauge 
fields by adding the effective potential $U(\bar{\Phi},\Phi,T)$ to the NJL 
Lagrangian~\cite{Ratti:2005jh, Mukherjee:2006hq, Roessner:2006xn, Costa:2008gr, Ghosh:2008et, Deb:2011en}. 
Further, in the fermion part of NJL model, covariant derivative substitutes the standard one. 
The PNJL Lagrangian is given by,

\begin{equation}
%\begin{split}
{\mathcal{L}}_{_{PNJL}} = \sum_f ({\bar{\Psi}}_f (i {\gamma_\nu} D^{\nu} - m_f) \Psi_f 
+ G_s [({\bar{\Psi}}_f \tau_a \Psi_f)^2 + (\bar{\Psi}_f i \gamma_5 \tau_a \Psi_f )^2])
+ {\it{U}} ( {\bar{\Phi}}, \Phi , T )
%\end{split}
\label{lagr1}
\end{equation}
Here $D_\nu$ is the covariant derivative, $D_\nu = \partial_\nu -i(g A_\nu + {\delta_0^\nu} \mu_f)$, 
$A_\nu = A_\nu^a {\tau^a \over 2}$. Here subscript $f$ refers to the $u$, $d$ quark flavors. This term 
takes into account the interaction between the gauge 
and quark fields. $m_f$ and $\mu_f$ are quark mass and chemical potential of quark flavor $f$ 
respectively. $G_s$ is the four quark contact interaction strength. The thermodynamic potential 
in the mean field approximation for the above theory with two quark ($u,d$) flavors is given 
by~\cite{Ghosh:2007wy, Ratti:2005jh},  
%{\bf Fukushima:2003fw, Sakai:2008um not citing here}
\begin{equation}
\begin{split}
\Omega = - \sum_{f= u,d} \int_0^{\infty} {{d^3 p } \over (2 \pi)^3}  ( 2T  {\rm ln} [ 1 + 3 \Phi e^{-\beta (E_f - \mu_f)} 
+ 3 {\bar {\Phi}} e^{-2 \beta (E_f - \mu_f)}  + e^{-3 \beta (E_f - \mu_f)}] \\
+ 2T  {\rm ln} [1+3 {\bar{\Phi}} e^{-\beta (E_f + \mu_f)} + 3 \Phi e^{-2 \beta (E_f + \mu_f)} + e^{-3 \beta (E_f + \mu_f)}]) \\
	- 6 \sum_{f= u,d} \int {{d^3 p } \over (2 \pi)^3} E_f ~ \theta(\Lambda -|{\vec p}|) + \sum_{f= u,d} G_s {\sigma_f ^2} + {\it{U}} ( {\bar{\Phi}}, \Phi , T ).
\end{split}
\label{omega1}
\end{equation}
Here $\sigma_f = \langle {\bar \Psi} \Psi \rangle_f $ is the quark condensate. $\mu_{u}$ 
and ${\mu_d}$ are $u$ and $d$ quark chemical potentials respectively. For the 
present calculations chemical potentials are set to $\mu_u = \mu_d = 0$. 
The masses of $u$ and $d$ quarks are taken to be degenerate, i.e $m_u=m_d=m_0$.  
$E_{u,d} = \sqrt{p^2 + \Sigma_{u,d}^2}$ are the single particle energies with 
$\Sigma_{u,d} = m_0 -G_s \sigma_{u,d}-G_s \sigma_{d,u}$ where $G_s=10.08$ GeV$^{-2}$ 
and $m_0=5$ MeV.  For $T_0=190$ MeV in the effective potential ${\it U}$,
the thermodynamic potential given by Eq.~{\ref{omega1}} results in qualitatively similar 
thermodynamic behaviour as that in~\cite{Ghosh:2007wy, Ratti:2005jh}.
%with temperature shows similar behaviour 
%~\cite{Costa:2010zw, Schaefer:2007pw, Torres-Rincon:2015rma}.

For the temperature dependence of the quark condensates and expectation value of the Polyakov loop, we 
minimize the thermodynamic potential $\Omega (\sigma,m,T_0,T)$ by numerically solving the following 
set of equations,
\begin{equation}
{\partial \Omega \over \partial \Phi_1} = 0, \quad {\partial \Omega \over \partial \Phi_2} = 0, 
\quad {\partial \Omega \over \partial \sigma} = 0.
\end{equation}

$\Phi_1$ and $\Phi_2$ are real and imaginary parts of $\Phi$. The numerical program requires initial 
trial values of $\Phi$ and $\sigma$. It evolves the trial values such that the thermodynamic potential 
decreases. The process stops once a minimum is reached within a certain numerical accuracy. This method 
can not find all the minima at once. For each minima the numerical procedure is repeated, by suitable 
choices of initial conditions.

One can show that, in the ground state $\Phi$ is real valued, i.e $\Phi_2=0$ ($\theta=0$). Hence, for 
the ground state, we solve the above equations with initial values of $|\Phi|>0$, and $\theta=0$.  
We took the zero temperature value of $\sigma$ as its initial value. For the meta-stable states, the $Z_3$ 
rotated values of the ground state $\Phi$ as initial value works well. Since the $Z_3$ symmetry is 
explicitly broken, $Z_3$ rotated $\Phi$ does not solve the equations. However, the meta-stable states
are found to be close to  $Z_3$ rotations of the stable state. 
The value of $\sigma$ in the meta-stable state is found to  
differ from the value in the stable state.

With above numerical procedure, in the temperature range of  $T_c(170)$ MeV - $T_m(194)$ MeV, only 
one solution is found with $\Phi_1>0$ and $\Phi_2=0$. $Z_3$ rotated values of this solution as 
initial condition does not result in any new solution. As the temperature is increased, for 
$T> 194$ MeV, two local minima appear around $\theta= {2\pi \over 3}$ and $\theta= -{2\pi \over 3}$. 
The thermodynamic potentials for these two states are found to be same, but higher than that for the
ground state for which $\theta=0$. The values of $\Phi$ for the meta-stable and stable states are 
no more related by $Z_3$ rotation. 

\begin{figure}[h]
  \centering
  \subfigure[]
  {\rotatebox{0}{\includegraphics[width=0.40\hsize]
      {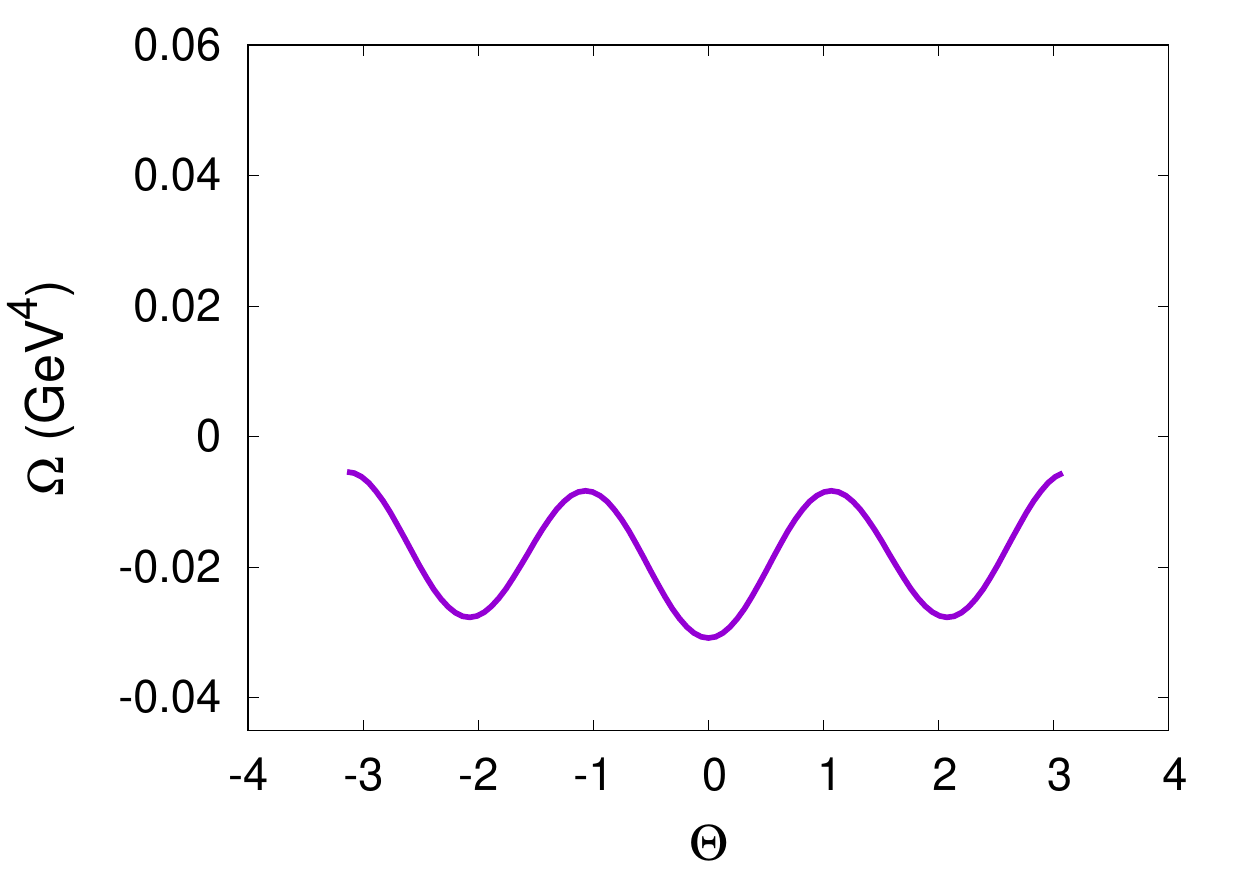}}
    \label{fig2a}
  }
  \subfigure[]
  {\rotatebox{0}{\includegraphics[width=0.40\hsize]
      {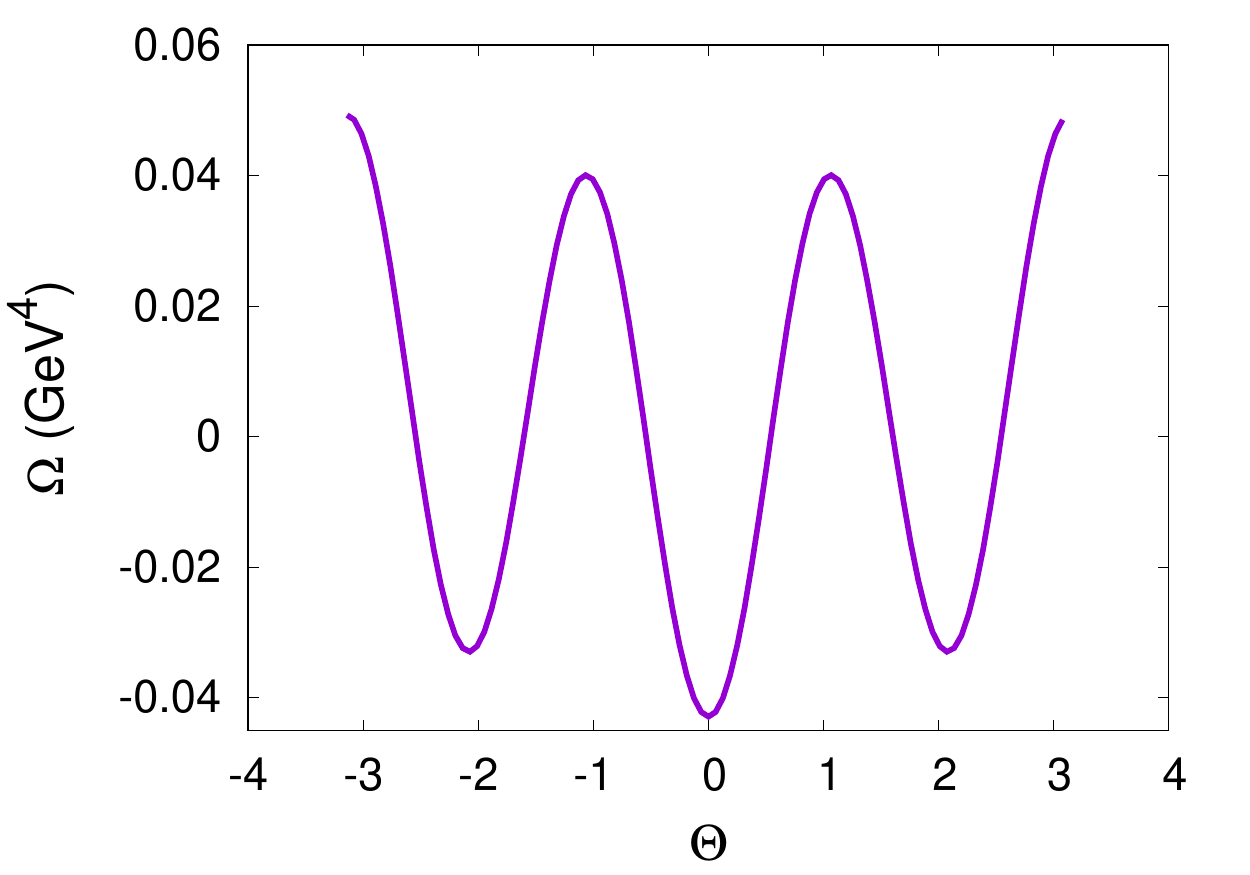}}
    \label{fig2b}
  }
\caption{Thermodynamic potential versus theta (a) at temperature 199.5 MeV
(b) at temperature 247 MeV. $\sigma$ value at each point is the one
which minimises $\Omega$ at the given value of $\theta$.}
\label{fig2}
\end{figure}

\begin{figure}[h]
  \centering
  \subfigure[]
  {\rotatebox{0}{\includegraphics[width=0.46\hsize]
      {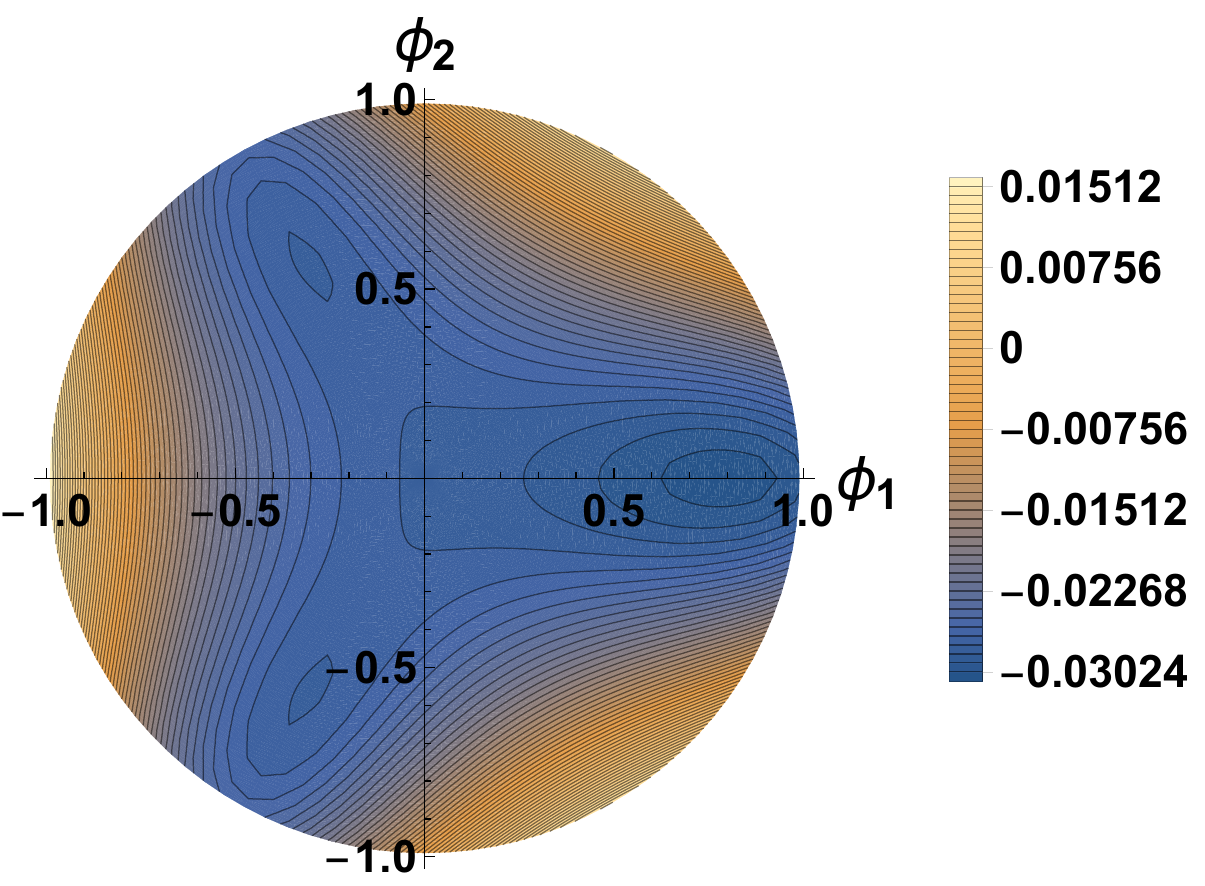}}
%    \label{cont1}
  }
  \subfigure[]
  {\rotatebox{0}{\includegraphics[width=0.46\hsize]
      {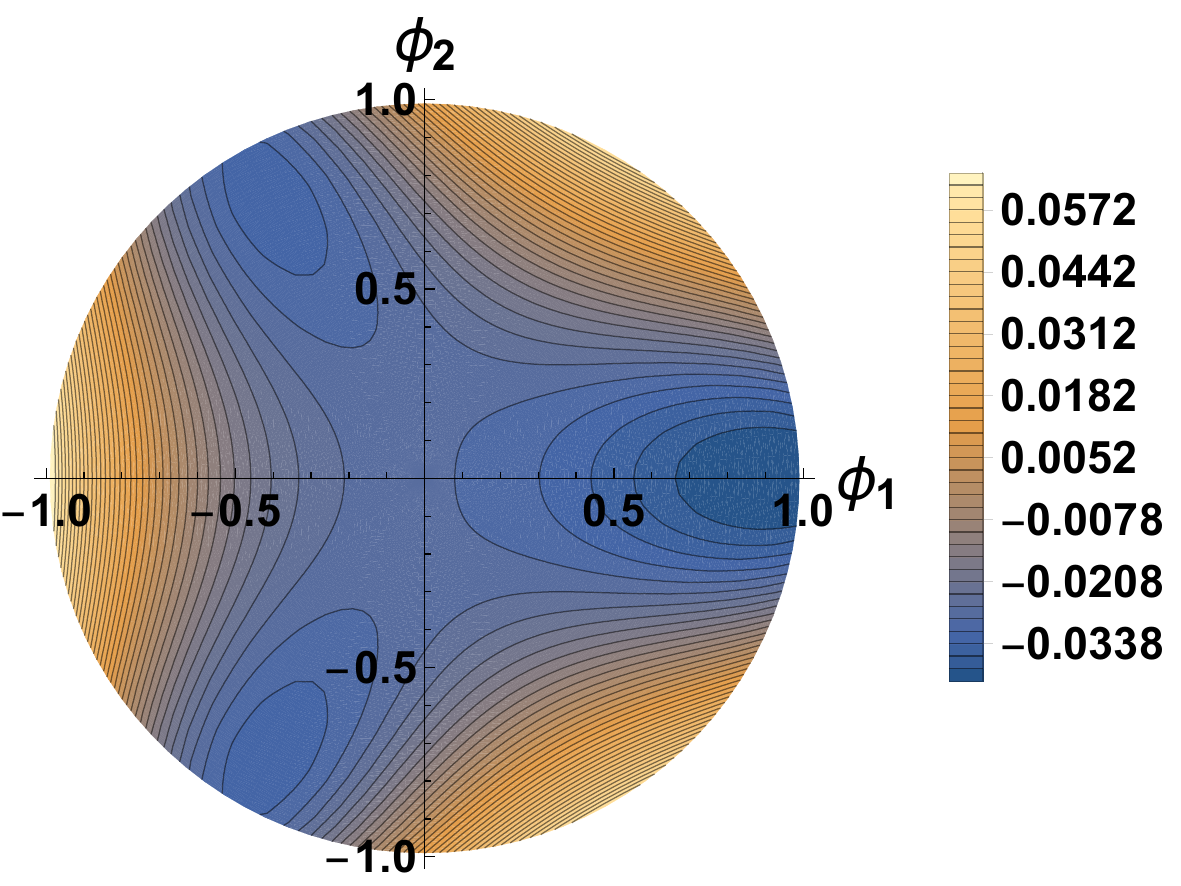}}
%    \label{cont2}
  }
%\label{fig3}
\caption{Contour plot of the thermodynamic potential on the $\Phi_1$ - $\Phi_2$ plane (a) 
at temperature 199.5 MeV (b) at temperature 247 MeV. $\sigma$ value at each point is the one
which minimises $\Omega$ at the given $\Phi_1$, $\Phi_2$. }
\label{fig3}
\end{figure}

In Fig.~\ref{fig2}, the thermodynamic potential versus $\theta$ has been plotted by fixing 
$|\Phi|$ in the stable state ($|\Phi| = 0.79763$ for $T=$199.5 MeV and  $|\Phi| = 0.92962$ for 247 MeV) and minimising thermodynamic potential with respect to $\sigma$. 
In Fig.~\ref{fig2a} the temperature is close 
to $T_m$ when the effective potential develops two saddle points. Comparing Fig.~\ref{fig2a} and 
Fig.~\ref{fig2b} one can clearly see signs that the barrier between meta-stable and stable states 
increases with temperature. In Fig.~\ref{fig3}, the contour plots of the thermodynamic potential 
in the $\Phi_1$-$\Phi_2$ plane are shown. At each point $\Phi_1$, $\Phi_2$ in the contour plot, we have 
fixed $\sigma$ at the value which minimises the thermodynamic potential.
For temperatures above $T_m$ there are two meta-stable states and one stable
state. We denote the meta-stable state for which $\Phi_2<0$ by M1 and the other
by M2. The stable state is denoted by SS. Subscripts $s$ and $ms$ on variables denote
their values in the stable state and the meta-stable states respectively.   

\begin{figure}[h]
  \centering
%  \subfigure[]
  {\rotatebox{270}{\includegraphics[width=0.50\hsize]
      {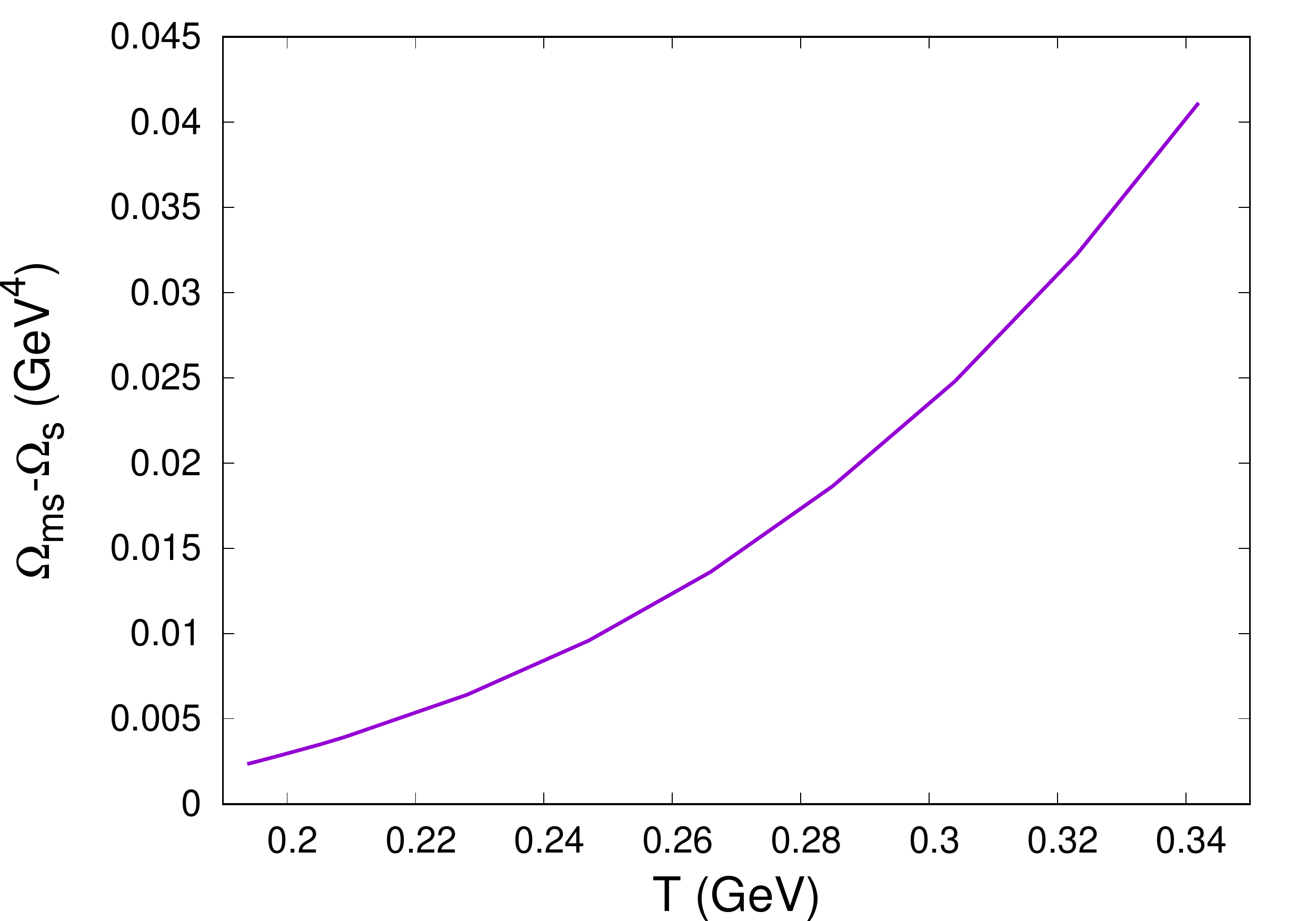}}
      }
\caption{ The thermodynamic potential difference between meta-stable and stable states vs. $T$.}
\label{omega}
\end{figure}

\begin{figure}[h]
  \centering
  \subfigure[]
  {\rotatebox{0}{\includegraphics[width=0.485\hsize]
      {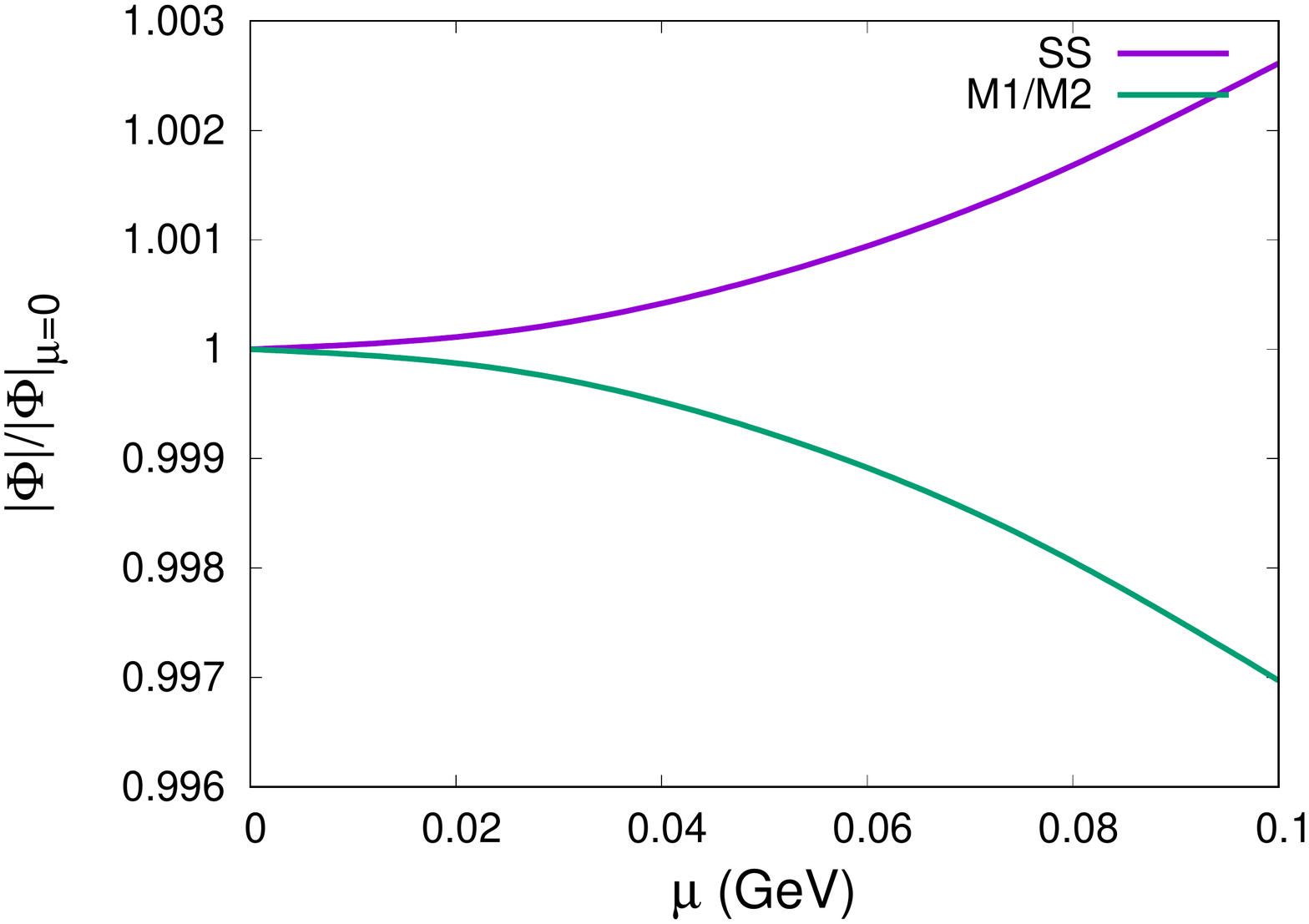}}
%    \label{modphi}
  }
    \subfigure[]
  {\rotatebox{0}{\includegraphics[width=0.485\hsize]
      {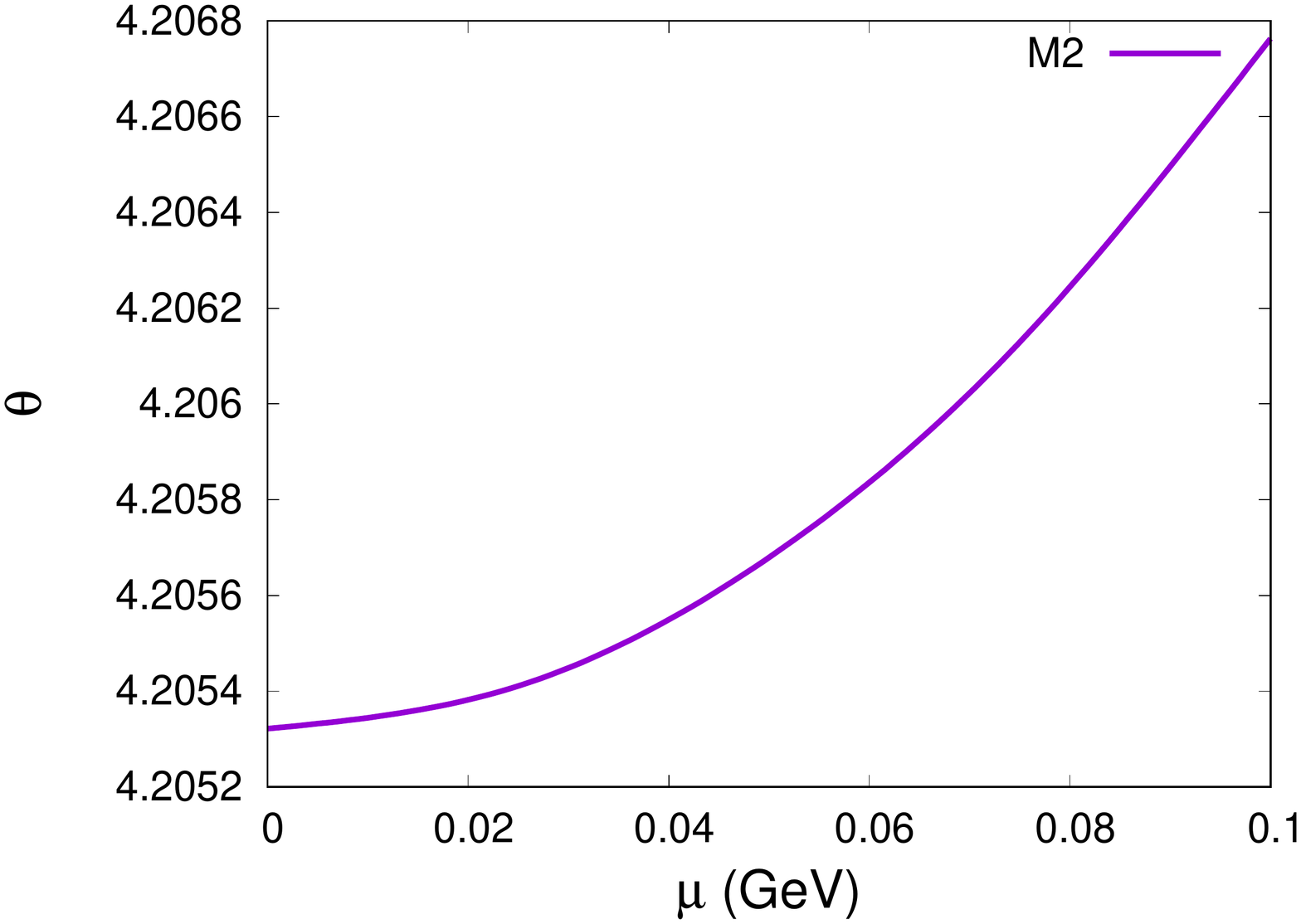}}
%    \label{phase}
  }
\caption{$\mu$ dependence of the absolute value of Polyakov loop in the stable and metastable states normalized to the corresponding values at $\mu=0$ (a) and the phase of the meta-stable state M2 (b) }
\label{modphi}
\end{figure}

Fig.~\ref{omega} shows the difference in the thermodynamic potential of the meta-stable to the 
stable state ($\Omega_{ms}-\Omega_{s}$) vs
$T$. This difference increases with temperature, which suggests enhancement in the explicit breaking
for larger temperatures.  In Fig.~\ref{modphi} we show the Polyakov loop in the M1, M2 and 
SS states for small values of $\mu$. We find that with increase in $\mu$ the absolute value of the Polyakov 
loop decreases in the meta-stable states while it increases in the stable state. $\mu$ contribution effectively
tilts the thermodynamic potential towards the stable state. We find that the phase of the M2 state shifts
toward higher value. The phase of M1 decreases as it is the complex conjugate
of M2. We also find that the barrier height between (M1, M2) and SS decreases while $T_m(\mu)$ increases slightly.

%\textcolor{red}{May be we should plot the barrier height as well?}.

%\subsection{High temperature behavior of thermodynamics states}


\section{Bounce solution for the decay of meta-stable states}

In PNJL model, with 2 quark flavors, meta-stable states exist above $T_m \sim 194$ MeV.
 Even though the value of $T_m$ in this model is too small compared to the lattice result
~\cite{Deka:2010bc}, we believe that it will give qualitative results for the effect of meta-stable states in heavy-ion collisions.
We mention here that, with other Polyakov loop potentials, higher $T_m$ can be achieved by tuning the 
integration cut-off in Eq. \ref{omega1} though the results do not qualitatively differ from the present case.

As mentioned before if a system is in the meta-stable state, it will eventually 
decay to the stable state. A meta-stable state can either
become unstable if the temperature drops below $T_m$ or decay via nucleation of bubbles like
in a first order phase transition. At finite temperature, there will be fluctuations in the 
form of bubbles, with stable states in their core. The free energy of a bubble
consists of two components, the volume component and the surface component. The volume component comes 
from the free energy difference between the stable and the meta-stable states. The surface component
comes from the fact that, the fields ($\Phi,\sigma$) have to interpolate between stable values
at the center to meta-stable values outside. For a critical bubble, these two components balance and
a small fluctuation can make it grow or collapse. Thus the critical bubble and its nucleation rate play 
an important role in a first order phase transition. For decay of the state M1(M2) the fields $\Phi$
and $\sigma$ will have values corresponding to the SS inside the bubble. Both these fields vary smoothly
across the bubble wall and approach the values corresponding to M1(M2). Given 
that the free energy does not depend on the sign  of $\Phi_2$, the bubbles interpolating M1 and SS will have same action as the other
interpolating M2 and SS. The critical bubble is obtained from 
the bounce solution which is a saddle point of the Euclidean action. We must mention here 
that the bubble nucleation picture here is not related to any phase transition 
but to the fact that the theory allows existence of meta-stable states above a certain 
temperature and they can tunnel into the stable state.

The decay rate of the false vacuum (meta-stable state) can be 
calculated in the semi classical approximation where the dominant contribution
comes from the configurations with the least action~\cite{Coleman:1977py, Callan:1977pt}, i.e bounce solutions. 
It is shown that such configurations, at zero temperatures have $O(4)$ 
symmetry, reducing the problem to one degree of freedom along the radial direction given by 
$r^2=|\bf x|^2+\tau^2$ in the Euclidean space. It has been shown that the
problem is equivalent to calculating the classical evolution 
of a particle in the Euclidean space in presence of the inverted potential $-V(\phi)$, where the particle rolls down from the stable vacuum 
and bounces up to the meta-stable one. The decay rate then can be 
written as the summation of all such ``bounces"~\cite{Coleman:1977py}. 
At high temperatures, owing to the periodicity of the field theory  in 
the ``time" direction, the field configurations will have $O(3)$ 
symmetry on a time slice~\cite{Linde:1981zj}. For a single scalar field theory with 
a meta-stable state, the bounce can then be calculated by solving the 
equation of motion, 
\begin{equation}
\frac{d^2\phi}{d r^2}+\frac{2}{r}\frac{d \phi}{d r}=\frac{\partial  V}{ \partial \phi}
\label{bnc1}
\end{equation} 
with the boundary conditions $\phi \rightarrow \phi^m$ as $r\rightarrow\infty$, where 
$\phi^m$ is the value of the field in the meta-stable state. For $r\sim 0$ the field is
expected to be close to the stable state.
If $r$ were the time variable, Eq.~\ref{bnc1} would be the equation of motion of a particle
in an inverted potential with a damping term. The required boundary conditions are 
equivalent to the trajectory of a particle 
starting from the maximum of the inverted potential (which corresponds to the stable state), 
rolling down and climbing up to the local maximum which corresponds to the meta-stable 
point. As the particle approaches the local maximum its velocity approaches zero. 
The critical bubble nucleation probability rate per unit volume
at finite temperature is proportional to $\exp(-S/T)$, where $S$ is the action of the
bounce solution.

\subsection{The bounce}

The bounce Eq.~{\ref{bnc1}} is non-linear in $\phi$, which makes it difficult to solve analytically. Only in
the thin-wall approximation, when the stable and meta-stable are almost degenerate, the bounce can be
calculated analytically. Such an approximation will not be valid in the present case as the difference in 
the thermodynamic potential between the stable and meta-stable states increases with $T$ and dominates 
the barrier height. Hence, numerical integration is the only way to find the bounce/bubble profile.  
The numerical integration is straight forward when there is a symmetry, for example in $U(1)$ theory, 
where only the radial mode of the field appears in the bounce equation. The phase is taken 
to be uniform, for minimum action bubble profile.  

In the PNJL model there is no such symmetry, the real and imaginary parts of Polyakov loop field 
and the sigma field are expected to have non-trivial profiles. Since evolving all the three fields 
simultaneously proved extremely difficult, we kept sigma field constant throughout the 
trajectory, that is, $\sigma$ is independent of $r$. Later we will consider sample profiles for $\sigma$ to 
estimate the corrections to the action. We also calculate the lower bound of the action.
%We have tried estimate the change in the action due to this approximation, which turns out
%to be below $10\%$. For all the scenarios of a $\sigma$ profile considered, the action always increases. 
In the present case, the thermodynamic potential $\Omega(\Phi_1,\Phi_2,\sigma)$ replaces $V(\phi)$ in Eq.  \ref{bnc1}.
The equations to be solved simultaneously are given by, 
\begin{eqnarray}
\nonumber
\frac{d^2\Phi_1}{dr^2}+\frac{2}{r}\frac{d\Phi_1}{dr}&=&\frac{\partial\Omega}{\partial\Phi_1} \\
\frac{d^2\Phi_2}{dr^2}+\frac{2}{r}\frac{d\Phi_2}{dr}&=&\frac{\partial\Omega}{\partial\Phi_2}
\label{phi2}
\end{eqnarray}
The boundary conditions are $\Phi_i \rightarrow \Phi_i^m$ as $r\rightarrow\infty$.  $\Phi_i^m$, $i=1,2$, are 
the values of $\Phi_1,\Phi_2$ in the meta-stable state. 
For the numerical integration $r$ is discretized as $r\rightarrow r_n=n\delta$, 
where $\delta$
is the lattice spacing.  $\delta$ must be small compared to the length scale of typical variations of $\Phi$. The 
integration starts from
$r=r_0\sim 0$. Two types of discretizations of Eq.~{\ref{phi2}} are considered. In the first approach, 
the values of $\Phi_1, \Phi_2$ at $r=r_0$ and $r=r_0+\delta$ are used to generate the trajectory. In the second approach 
the two equations are rewritten as four first order equations. In this case, the values of $\Phi_1, \Phi_2$ as well as their 
derivatives at $r_0$ determine the trajectory. It has been checked that both these methods of integration give same results. 
A few other approaches of calculating bounce solution for multiple field cases are discussed in~\cite{Konstandin:2006nd, 
Wainwright:2011kj, Piscopo:2019txs, Masoumi:2016wot, Athron:2019nbd, Sato:2019axv, Profumo:2010kp, Chigusa:2019wxb}.

\begin{figure}[h]
  \centering
  {\rotatebox{0}{\includegraphics[width=0.55\hsize]
      {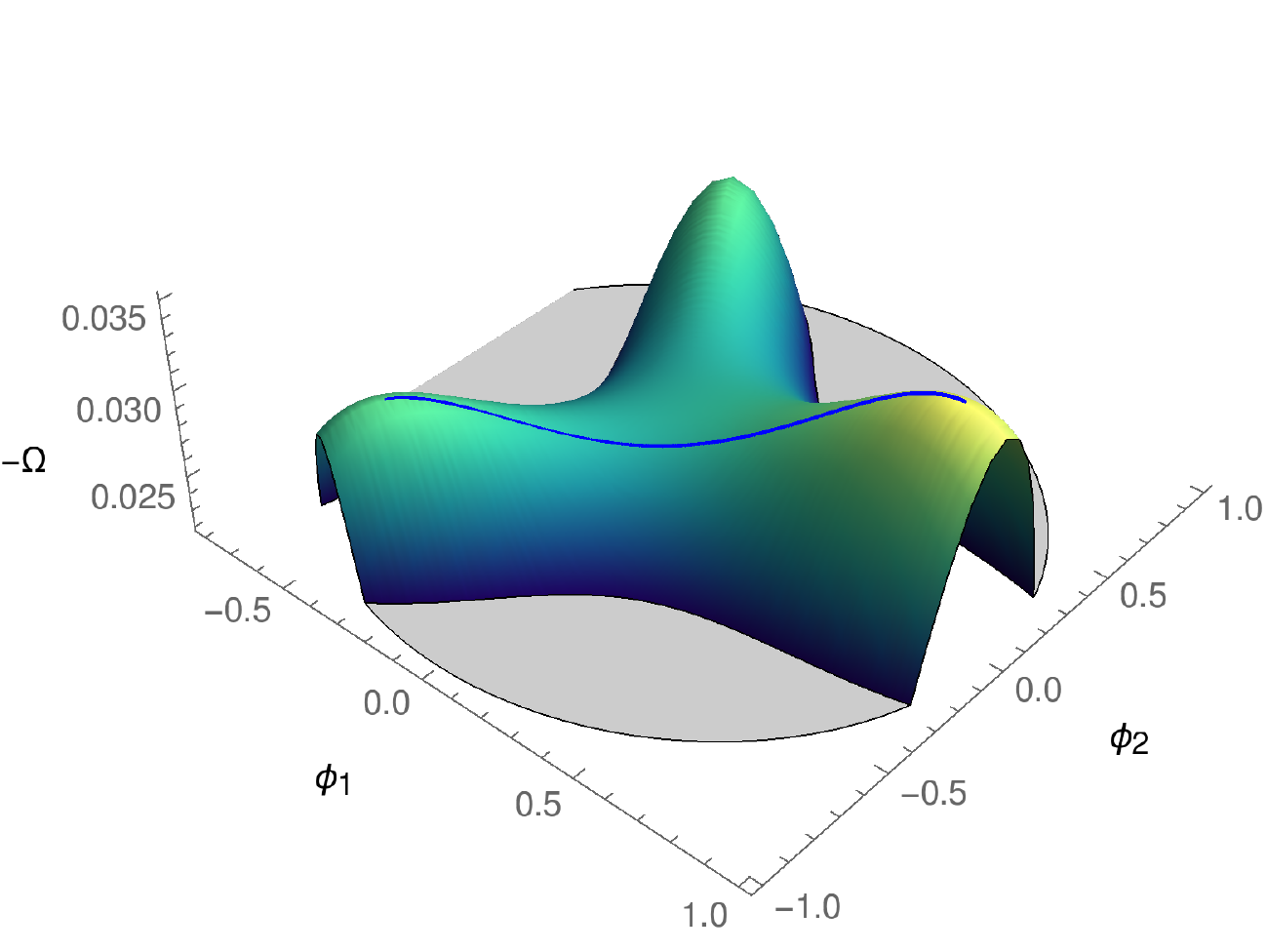}}
  }
\caption{Plot of the inverted potential in the Polyakov loop plane. The blue
trajectory is the bounce solution at 228 MeV.}
    \label{ridge}
\end{figure}

In Fig.{~\ref{ridge}} we show the plot of the inverted potential in the Polyakov loop plane.
 There is a ridge which connects the stable and meta-stable
states. The height of the ridge initially drops from the stable peak but eventually rises 
to the meta-stable peak. The bounce profile must start near the stable state and 
approach the meta-stable state. This can happen only for a unique choice of initial conditions, i.e.
position $(\Phi_1, \Phi_2)$ and velocity $(\frac{d\Phi_1}{dr},\frac{d\Phi_2}{dr})$. For wrong choices,
 the trajectory will fall off to infinity either through the center along $\theta = \pi$
or by crossing the ridge to $\theta \sim -\pi/3$. Hence, the initial conditions must be tuned which is
achieved by standard bisection method. The basic idea is that with the given initial conditions,
position ($\Phi_1(r_0),\Phi_2(r_0))$ and velocity (${d\Phi_1 \over dr}|_{r_0},{d\Phi_2 \over dr}|_{r_0}$), 
if a trajectory undershoots  (overshoots) the meta-stable point (peak), we
start with an initial choice (position) closer to (farther from) the ``global'' minimum (global peak).
Further the bisection method is used to find the direction of initial velocity such that the undershoot
trajectory makes a $180^o$ turn or the overshoot trajectory passes through the meta-stable state
peak. We  have
checked that the results do not change for smaller $\delta$. 
%\textcolor{red}{no need to give details}.

\subsection{The bubble}
\begin{figure}[h]
  \centering
  {\rotatebox{-90}{\includegraphics[width=0.50\hsize]
      {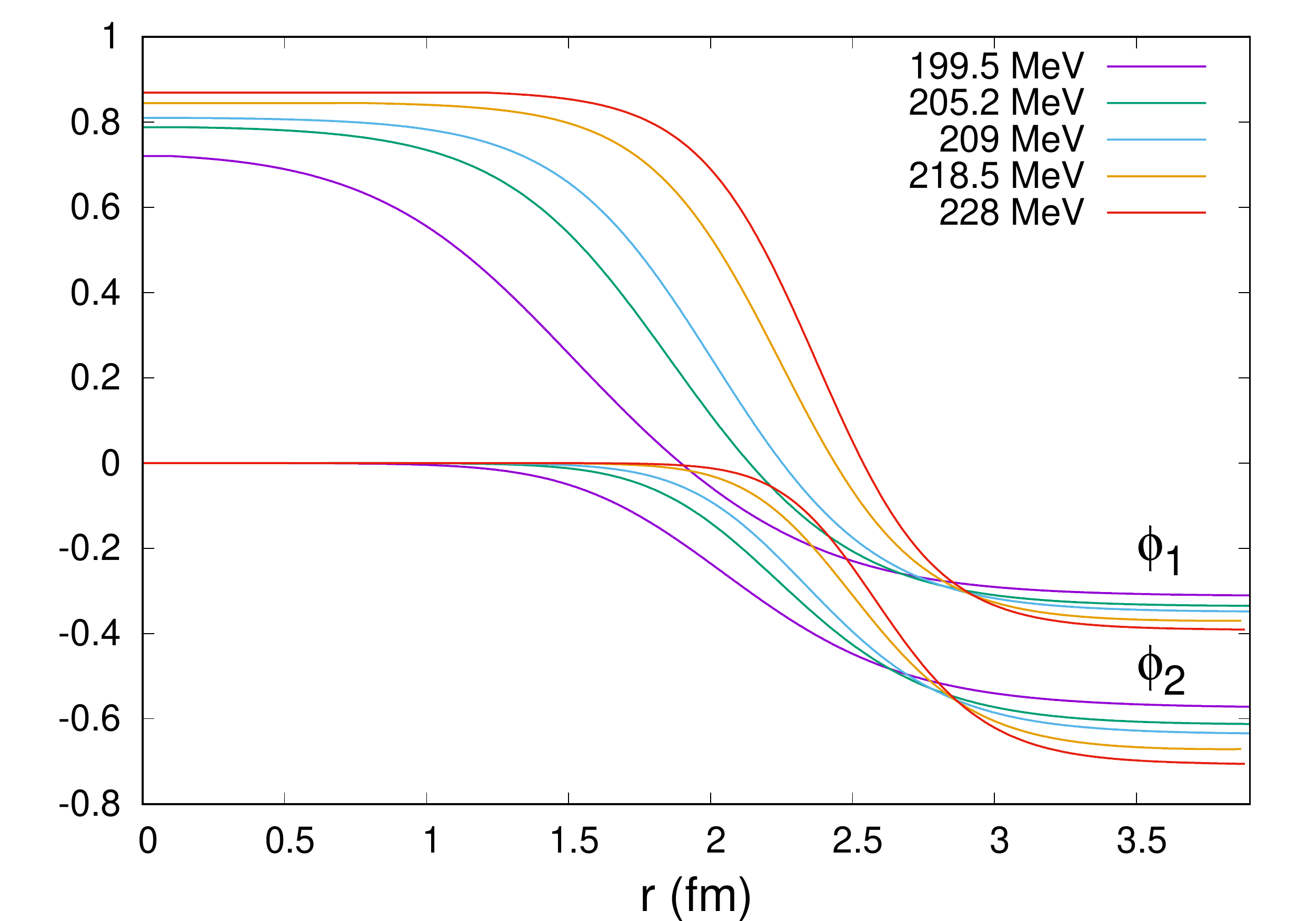}}
  }
\caption{Bubble profiles for $\Phi_1$ and $\Phi_2$ at different temperatures.}
 \label{bbls}
\end{figure}

The bubble profiles are computed for temperatures in the range $T=1.05 T_0~-~1.2 T_0$, i.e., 199.5 MeV to 228 MeV.
Fig.~\ref{bbls} shows the temperature dependence of the bubble profiles. These bubble profiles
represent the decay of M1 to SS. The values of $\Phi_1$ 
and $\Phi_2$ approach asymptotically to their corresponding meta-stable values. For
 temperatures just above $T_m$ the barrier between the stable and meta-stable states is 
small compared to $\Omega_s-\Omega_{ms}$. Starting with initial values of the field close to 
$\Phi_i^s$ at $r_0\sim 0$ will always lead to overshooting. Hence the initial values of the fields (at the center of the bubble) must be farther away from the stable point. Since the field starts already on a higher slope for small $r$, damping dominates the profile giving a broad ``wall" profile for the bounce. For higher temperatures, the initial point is 
closer to the stable point. The force term is small; so is acceleration. The field gets to
 spend more time near the stable state. Therefore the core radius of the bubble increases as we go towards higher temperatures. The bubble ``wall" is thinner because the particle/field spends 
larger time near the stable maximum and when it eventually starts rolling, the damping is small. 

\begin{figure}[h]
  \centering
  {\rotatebox{-90}{\includegraphics[width=0.50\hsize]
      {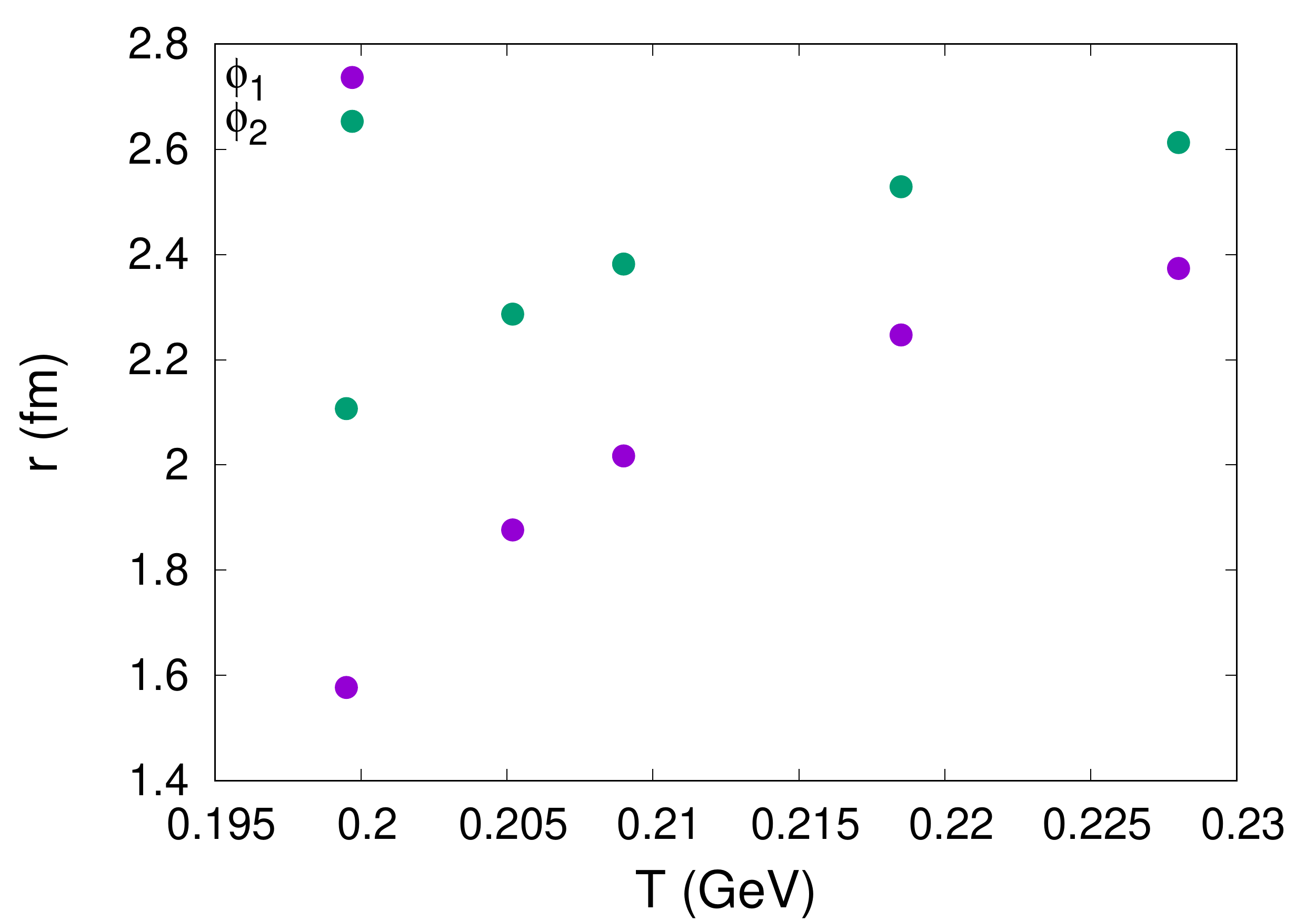}}
  }
\caption{Bubble radii vs temperature}
    \label{rad}
\end{figure}

Fig.~\ref{rad} shows the radii of these bubbles as a function of temperature. 
We define the radius of the bubble as the radial distance from the center to 
the point where the field drops half way to the meta-stable value.
%radius where the field value is the average of the value at the 
%core and that at infinity. 
Here we notice that the radii for the two different fields are not 
the same. The radius of the $\Phi_2$ profile is slightly higher than that of 
 $\Phi_1$. This is because the curvature of the potential along 
$\Phi_1$ and $\Phi_2$, or in other words, their mass scales, are different.

\begin{figure}
\centering
%\subcaptionbox{\label{vec}}{%
\includegraphics[origin=c,height=6.2cm]{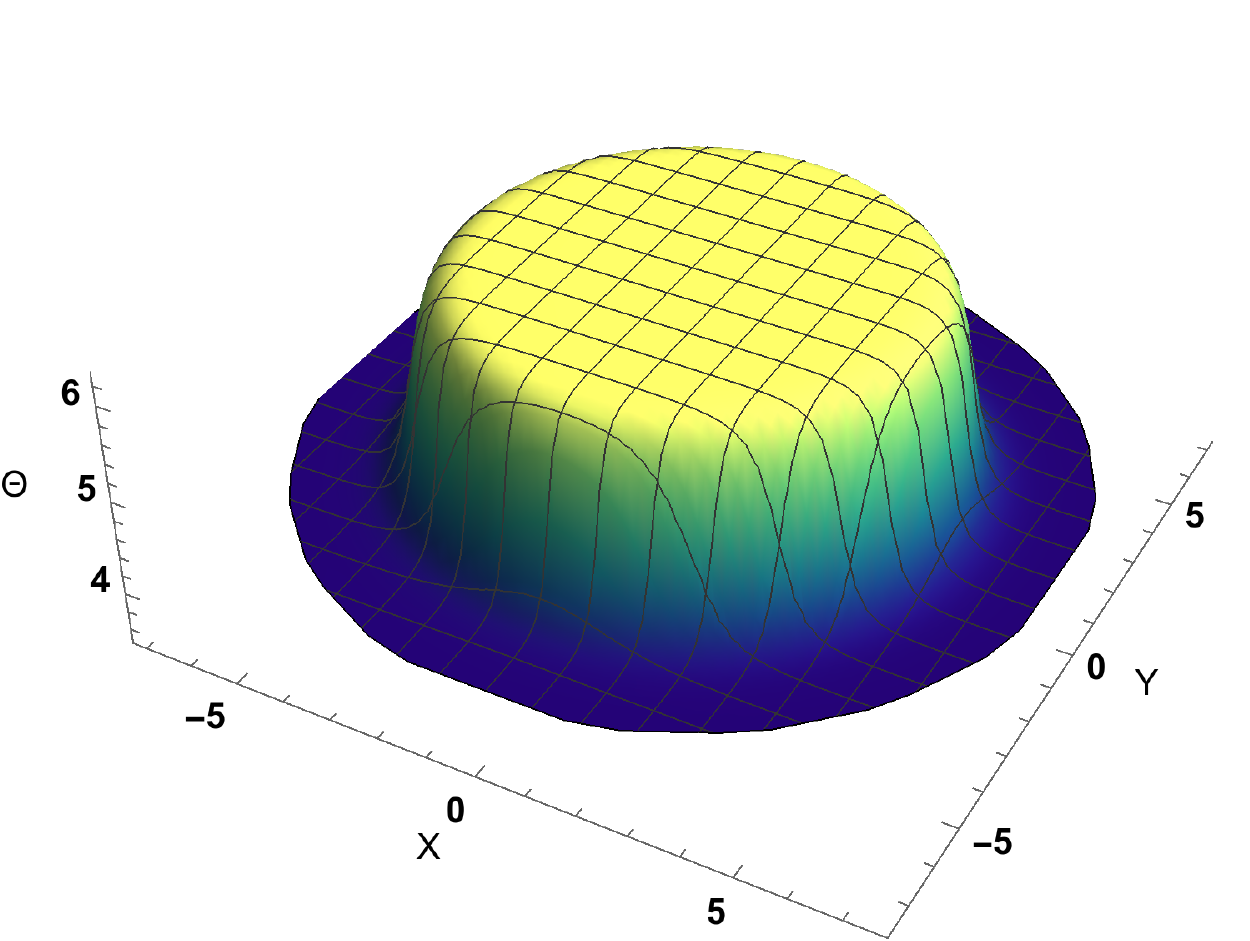}
\quad
%\subcaptionbox{\label{mod}}{%
\includegraphics[angle=-90,origin=c,height=5.7cm]{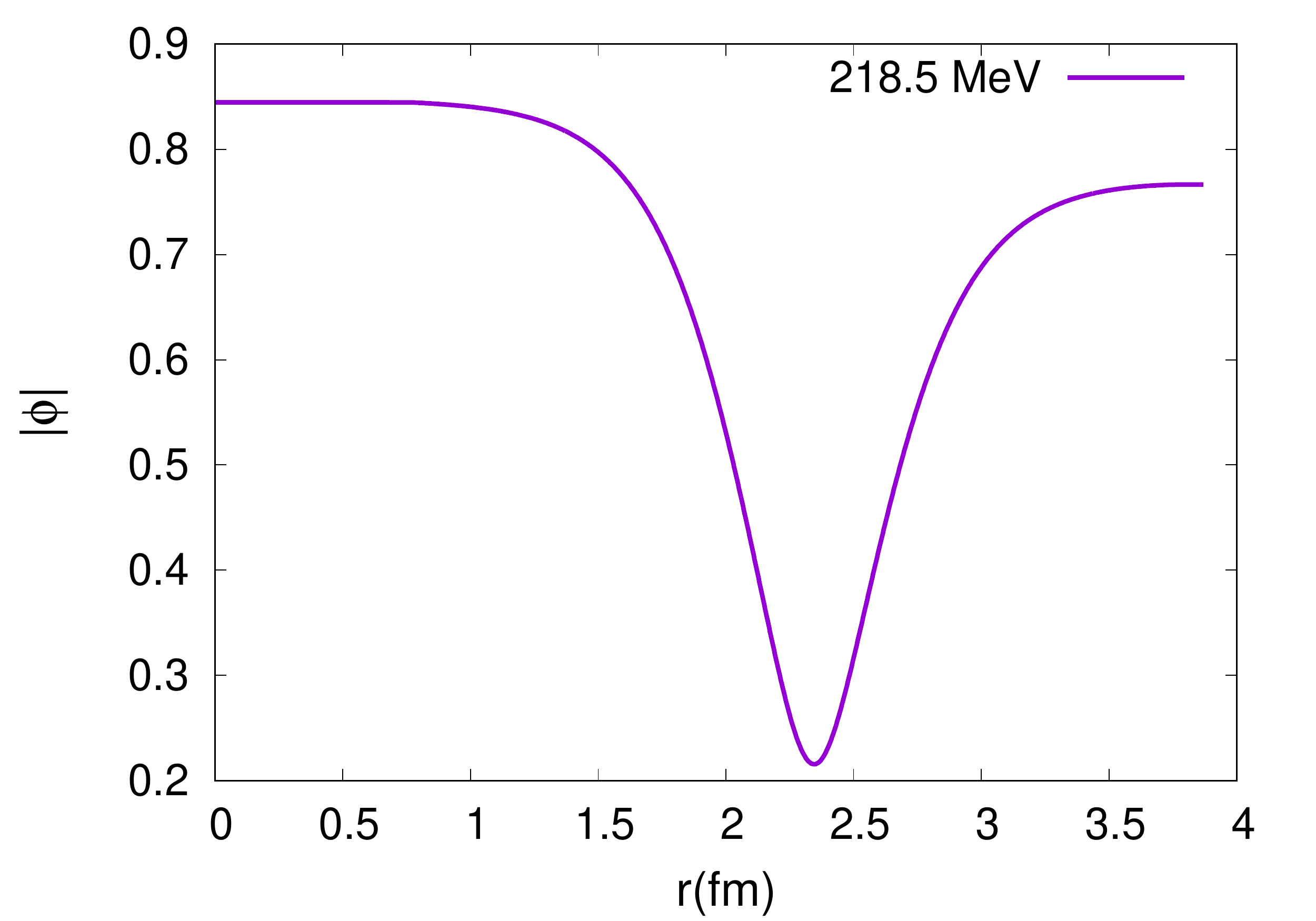}
\caption{Plot of $\theta$ field for the bubble (Left) and modulus of $\Phi$ vs. radius 
(Right) at 218.5 MeV.}
\label{mod1}
\end{figure}

Fig.~\ref{mod1} (left), shows a plot of the $\theta$ profile of the bubble at $T=218.5$ MeV. 
We also plot the magnitude of the Polyakov loop versus radius in Fig.~\ref{mod1} (right).

\section{Evolution of meta-stable state in heavy-ion collisions}

\begin{figure}[h]
  \centering
  \subfigure[]
  {\rotatebox{-90}{\includegraphics[width=0.34\hsize]
      {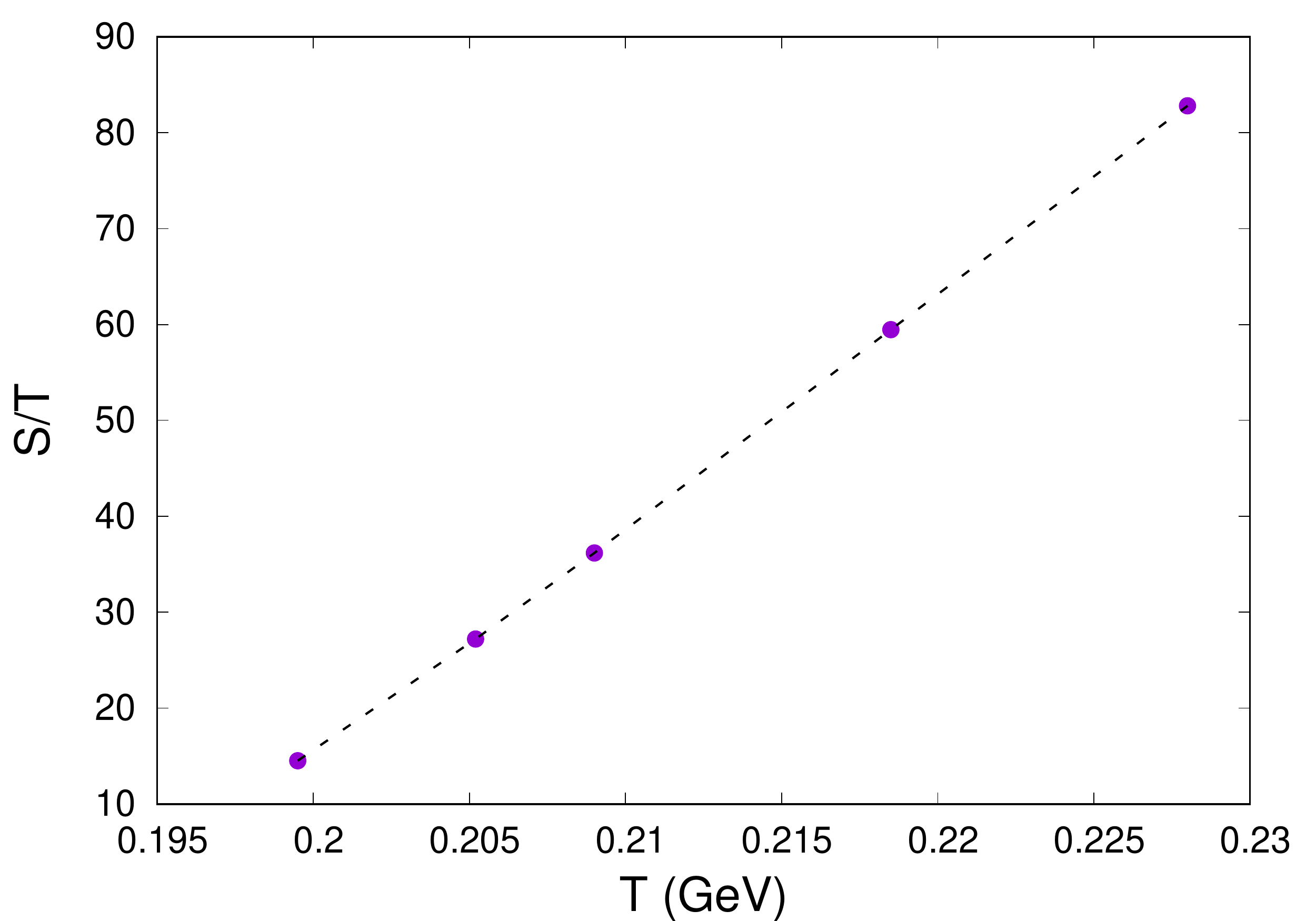}}
    \label{action}
  }
  \subfigure[]
  {\rotatebox{-90}{\includegraphics[width=0.34\hsize]
      {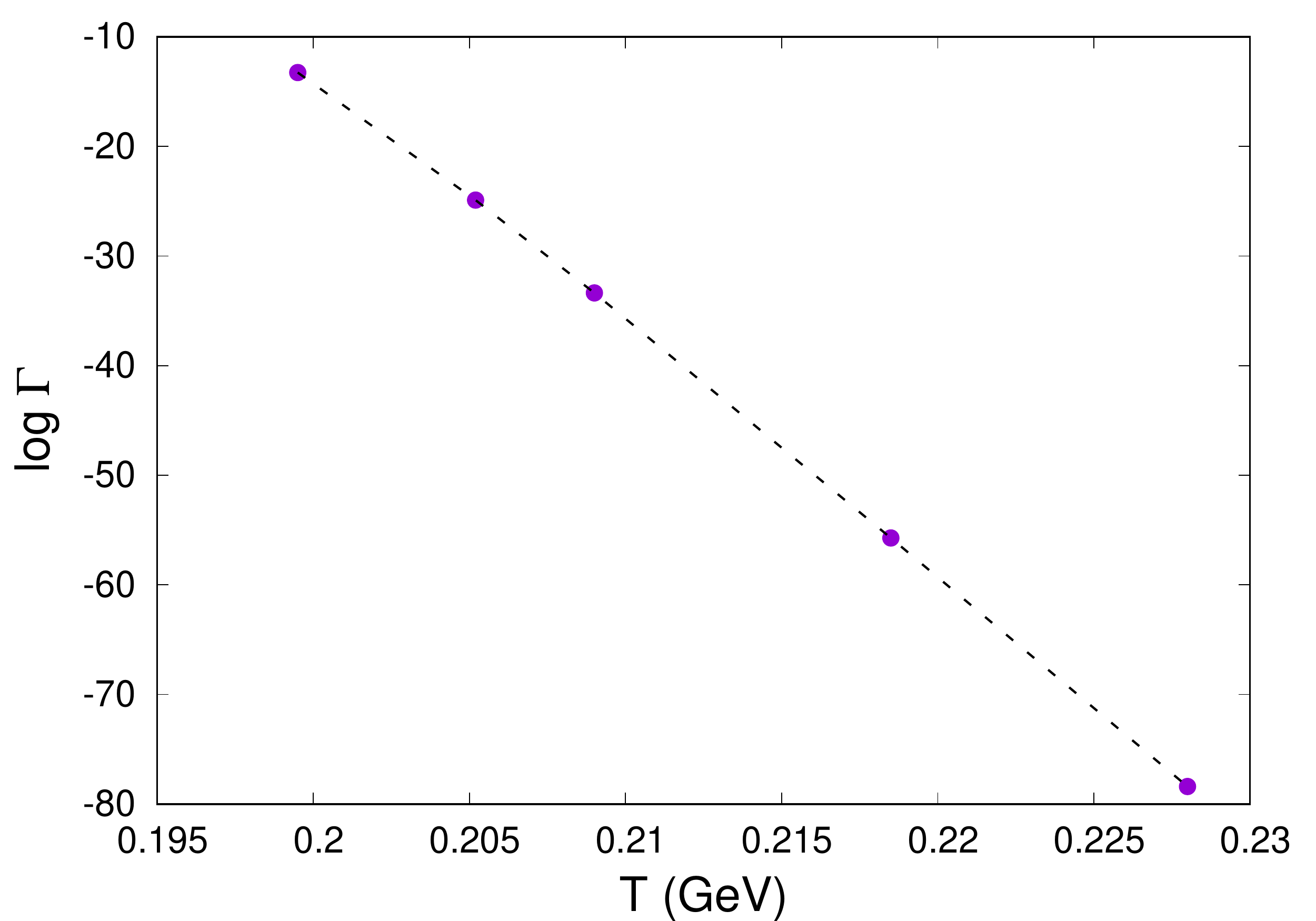}}
    \label{decay}
  }
\caption{(a) Bubble action in units of $T$ and (b) Log of the bubble nucleation rate (per $fm^3$ per 
$fm$ time) vs . $T$.}
\end{figure}

The decay rate of these meta-stable states depends on the bubble action, which is given by,
\begin{equation}
S=\int{4\pi r^2 dr \Big[\frac{1}{2}\alpha T^2\left\{\left(\frac{d\Phi_1}{dr}\right)^2
+\left(\frac{d\Phi_2}{dr}\right)^2\right\}+
\frac{1}{2}G_s^2\left(\frac{d\sigma}{dr}\right)^2+\Omega(\Phi_1,\Phi_2,\sigma)\Big] }
\end{equation}
\noindent
Here $\alpha$ is a constant given by $2N/g^2$~\cite{Dumitru:2000in}, where $N$ is the number 
of colors and $g$ is the gauge coupling constant. For $g/4\pi = 0.3$, $\alpha = 1.6$. 
Fig.~\ref{action} shows the plot of the bubble action in units of temperature vs $T$.
Let us recall here that $\sigma$ was kept constant at the meta-stable value in the bubble. For an estimate of the change in the action, an approximate $\sigma$ profile is computed by minimizing $\Omega$ with respect to $\sigma$ for a given $\Phi_1, \Phi_2$ profile. 
With the $\sigma$ profile the bubble action decreases slightly. The decrement is below $10\%$ for all the temperatures calculated. We also checked with $\sigma$ profiles scaled like 
both $\Phi_1$ and $\Phi_2$ profiles, interpolating between $\sigma_s$-$\sigma_{ms}$. In the case where $\sigma$ profile was scaled like $\Phi_1$
the action was minimum (less than $20\%$ decrement). 

\vskip 0.2cm
The bubble nucleation probability per unit time per unit volume, or in  other words, the decay 
rate of the meta-stable state is given by \cite{Linde:1981zj}
\begin{equation}
\Gamma=T^4\Big(\frac{S}{2\pi T}\Big)^{3/2}\exp(-S/T)
\label{gma}
\end{equation}  

One can see from Fig. \ref{decay} that this value is as small as $10^{-6}$/fm
for the smallest temperature above $T_m$ (199.5 MeV) and grows insignificant 
for higher temperatures. Though it is difficult to solve for the bubble in the full
case by including equation corresponding to $\sigma$ in Eq.~\ref{phi2}, we can compute the
lower bound of the full action. The corresponding nucleation rate will then be
the upper bound. Note that $\Gamma$ has a peak in $S/T$. However, for the range of 
 $S/T$ in our calculation, $\Gamma$ is a monotonically decreasing function
of $S/T$.  To compute the upper bound on the nucleation rate we fix the $\sigma$ field
at the SS value ($\sigma_s$). To see how $\sigma(r)=\sigma_s$ leads to this bound,
we write the free energy ($F_b$) of a critical bubble of radius $R_d$ 
as 
\begin{equation}
F_b = -{4\pi \over 3} \rho R_d^3 + 4\pi\delta R_d^2,~~~~R_d={2\delta \over \rho}.
\end{equation}
where $\rho$ is the free energy difference between the stable and meta-stable
state, $\delta$ is the free energy cost (surface tension)  as the fields vary smoothly 
between the two states. The position of the meta-stable state when the $\sigma$ is fixed 
at $\sigma_s$ deviates from the full case such that $|\Phi_s - \Phi_{ms}|$ decreases. 
This effectively leads to decrease in $\delta$.  Also with $\sigma=\sigma_s$
the meta-stable states always have higher $\Omega_{ms}$ compared to the full case. This
leads to increase in $\rho$. Hence the free energy for $\sigma=\sigma_s$ is lower
compared to the full case. Note that the barrier height also plays a role in determining
$\delta$ which is the reason the bubble action grows with temperature. However for
a given temperature the barrier would slightly decrease as $\sigma$ changes from 
$\sigma_{ms}$ to $\sigma_s$.

\vskip 0.2cm

\begin{figure}[h]
  \centering
  \subfigure[]
  {\rotatebox{0}{\includegraphics[width=0.48\hsize]
      {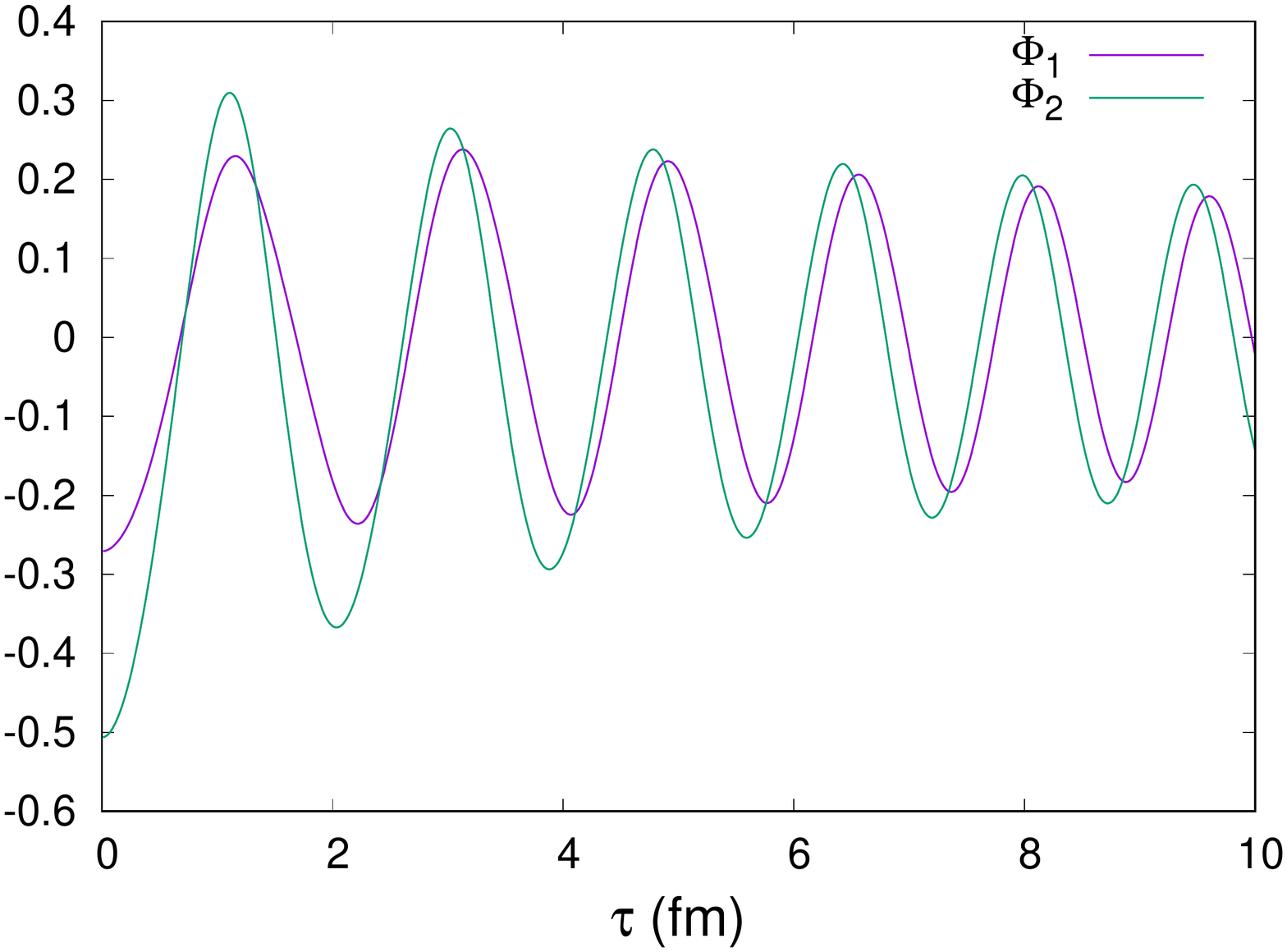}}
    \label{evl1}
  }
  \subfigure[]
  {\rotatebox{0}{\includegraphics[width=0.48\hsize]
      {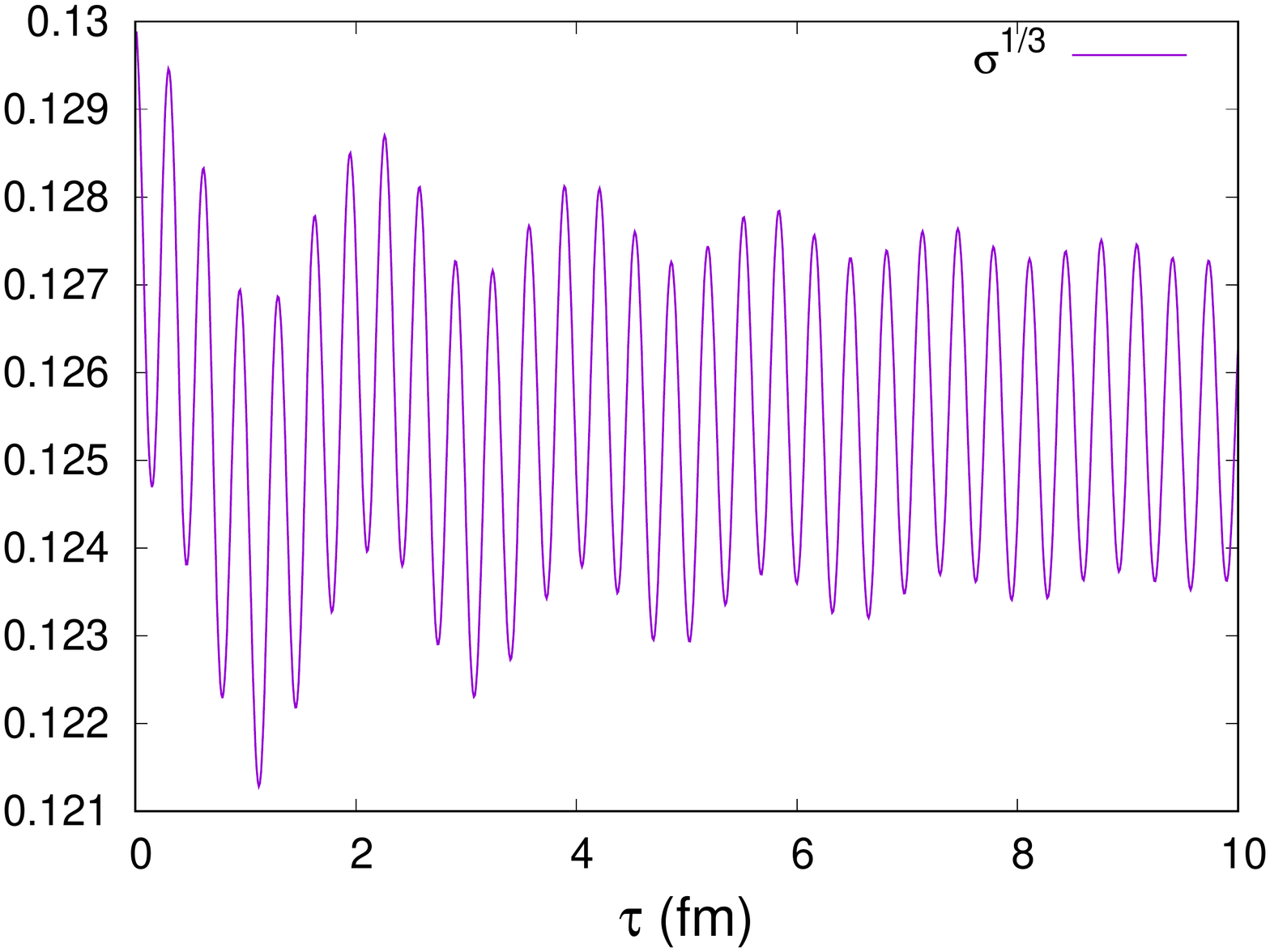}}
    \label{evl2}
  }
\caption{Evolution of (a) Polyakov loop and (b) $\sigma^{1/3}$  
after the temperature falls below $T_m$. Here, $\tau = 0$ corresponds to the time at 
which temperature is below
$T_m$.}
\label{evl}
\end{figure}

%All this indicates that if the system falls into one of the meta-stable 
%states during thermalization at high temperature, the state likely
%will not decay through bubble nucleation \textcolor{red}{Why?}. 
%This will have interesting consequences in heavy-ion collisions.
We do a quick calculation of the decay rate for the case of heavy-ion collisions, assuming
the thermalization time to be $\tau_i = 0.38$ fm and the initial temperature
of the order of $T_i \sim 550$ MeV. We consider the QGP to be cylindrical (the midrapidity region) with radius $8$ fm and
length $3$ fm.  The number of bubbles nucleated within this volume when the system cools down to
a temperature $T$ is estimated as follows. The bubble action $S$ as a function of temperature is obtained 
by fitting our data points. We used a longitudinally boost invariant $2+1 D$ 
hydrodynamic simulation with Glauber optical 
initial conditions following \cite{Kolb:2000sd}, to fit the temperature evolution. The number of 
bubbles nucleated in volume $V$ during the time $\tau$ when the temperature drops to $T$ is given by

\begin{equation}
N(\tau) = V\int_{\tau_i}^{\tau} \Gamma(t) dt.
\end{equation}

We find that $N(T_m) = N(\tau_m)$, where $\tau_m$ is the time at which 
temperature is $T_m$, is vanishingly small. Here we have used the profile with
sigma scaled as $\Phi_1$. If any of the other profiles are used, this number
only decreases.
The value of $N(\tau)$ rapidly decreases for higher temperatures.
If we assume a larger equilibration time, that is, a smaller
initial temperature, the value of $N(\tau)$ decreases further.
We have also computed the free energy of the bubble (action) and the nucleation rate 
by fixing $\sigma=\sigma_s$. We find that the bubble action is smaller 
by an average factor of $\sim 0.45$ compared 
to the case when $\sigma$ is fixed at $\sigma_{ms}$. In this calculation
the nucleation rate increases by a factor of hundred, though still remaining negligibly
small. We have considered the effects of $\mu$ upto $\sim 100$ MeV.
For $\mu=100$ MeV $T_m$ increases  by $\sim 2$ MeV. Since the barrier height decreases
with $\mu$, the bubble action $ S(T,\mu) < S(T,\mu=0)$. 
However the system spends lesser time above $T_m(\mu)$, hence the 
results will not change qualitatively.  
Hence, we do not expect any bubble nucleation in heavy ion collisions.
This leads to an interesting 
scenario, known as the spinodal decomposition. When the meta-stable states 
become unstable below $T_m$, the field will roll down to the minimum resulting 
in large angular fluctuations. Fig. \ref{evl} shows the evolution of the Polyakov loop and
chiral condensate at the centre of the quark gluon plasma after the 
temperature falls below $T_m$ at zero chemical potential. Since the meta-stable states
become unstable, the fields roll down and oscillate around the 
minimum. These oscillations will have interesting consequences in the dynamics of heavy-ion 
collisions including flow, jet energy loss~\cite{Lin:2013efa} and also may possibly 
lead to coherent emission of particles. As discussed above, at a small finite chemical
potential, $T_m(\mu)$ increases, and spinodal decomposition is expected to 
occur earlier.

%In the case of early universe since the system starts thermalised at a very high temperature, 
%the probability of regions falling into meta-stable states is very high. In this case there 
%will be domain walls stretching between the vacua. We have seen that the 
%probability of decay of these states is very small and hence these domain walls
%will survive close to $T_m$. Once these states become unstable, there 
%will be regions rolling down to the true vacuum and this may have consequences
%like particle production and local reheating.  

\section{Conclusions}
We have studied the $Z_3$ meta-stable states in PNJL model. The meta-stable states exist at and 
above the temperature $T_m \sim 194$ MeV. For small values of $\mu$ we found that $T_m$ increases slightly, i.e
by $\sim 2$ MeV at $\mu=100$ MeV.  The barrier height between the meta-stable and stable states decreases.
We {have discussed} the probability of the decay of these meta-stable states by 
calculating the stable bubble nucleation probability in the meta-stable regions using bounce 
solution. The bubble action measured in the units of temperature increases roughly linear in temperature. For small $\mu$,
relevant for heavy-ion collisions, the bubble action decreases slightly though the system spends lesser time
above $T_m$. Our results suggest that, the probability of these 
states decaying by tunnelling into stable states is very small in the case of heavy-ion collisions. Ultimately the meta-stable state will become unstable and the fields will start rolling towards the minimum.  This will lead to large oscillations of the Polyakov loop field, which may have interesting consequences to the dynamics of flow, jet energy loss, and also may lead to coherent emission of particles.

%We show three degenerate vacua correspond to $Z_3$ spontaneous symmetry
%breaking in pure $SU(N)$ gauge theory in Polyakov loop effective potential.
%We have studied the behavior of these vacua when dynamical quarks are included.
%In presence of quarks, the $Z_3$ symmetry is explicitly broken as a result two of 
%these vacua are sifted and they appear above transition temperature. 
%In particular these vacua exist above temperature {\bf 378 MeV}. These 
%vacua are the meta-stable states as they have lesser pressure and energy 
%values than the stable state. We also show that the depth of the meta-stable states 
%increases with increase in temperature. It would be interesting to study the behavior 
%of these meta-stable states at high temperatures.  
\acknowledgments

We thank A. P. Balachandran, Shreyansh S. Dave, Ajit Srivastava, Rajarshi Ray and
Ranjita Mohapatra for important comments and suggestions.

\end{document}